\documentclass[]{aastex631}

\usepackage{comment}
\usepackage{hyperref}
\usepackage{amsmath}
\usepackage{cleveref}
\usepackage{enumitem}
\usepackage{graphicx}
\usepackage{lmodern}

\shorttitle{Multi-Epoch Images of RW~Cep Following the Great Dimming }
\shortauthors{Anugu et al.}
\graphicspath{{RW_Cep_paper}{Figures/}}

\begin{document}

\title{Time-Evolution Images of the Hypergiant RW~Cephei During the Re-brightening Phase Following the Great Dimming}

\correspondingauthor{Narsireddy Anugu} \email{nanugu@gsu.edu}

\author[0000-0002-2208-6541]{Narsireddy Anugu}
\affiliation{The CHARA Array of Georgia State University, Mount Wilson Observatory, Mount Wilson, CA 91023, USA}

\author[0000-0001-8537-3583]{Douglas R. Gies}
\affiliation{Center for High Angular Resolution Astronomy and Department  of Physics and Astronomy, Georgia State University, P.O. Box 5060, Atlanta, GA 30302-5060, USA}

\author[0000-0002-9288-3482]{Rachael M. Roettenbacher}
\affiliation{Department of Astronomy, University of Michigan, Ann Arbor, MI 48109, USA}

\author[0000-0002-3380-3307]{John D. Monnier}
\affiliation{Department of Astronomy, University of Michigan, Ann Arbor, MI 48109, USA}

\author[0000-0002-7540-999X]{Miguel Montarg\'es}
\affiliation{LESIA, Observatoire de Paris, Universit\'e PSL, CNRS, Sorbonne Universit\'e, Université Paris Cit\'e, 5 place Jules Janssen, 92195 Meudon, France}

\author[0000-0003-2125-0183]{Antoine M\'erand}
\affiliation{European Southern Observatory, Karl-Schwarzschild-Str. 2, 85748 Garching, Germany}

\author[0000-0002-8376-8941]{Fabien Baron}
\affiliation{Center for High Angular Resolution Astronomy and Department  of Physics and Astronomy, Georgia State University, P.O. Box 5060, Atlanta, GA 30302-5060, USA}

\author[0000-0001-5415-9189]{Gail H. Schaefer}
\affiliation{The CHARA Array of Georgia State University, Mount Wilson Observatory, Mount Wilson, CA 91023, USA}

\author[0000-0003-2075-5227]{Katherine A. Shepard}
\affiliation{Center for High Angular Resolution Astronomy and  Department of Physics and Astronomy, Georgia State University,  P.O. Box 5060, Atlanta, GA 30302-5060, USA} 

\author[0000-0001-6017-8773]{Stefan Kraus}
\affiliation{Astrophysics Group, Department of Physics \& Astronomy, University of Exeter, Stocker Road, Exeter, EX4 4QL, UK}

\author[0000-0002-9759-038X]{Matthew D. Anderson}
\affiliation{Georgia Tech Research Institute, Baker Building, 925 Dalney St. Atlanta, GA 30318}

\author{Isabelle Codron}
\affiliation{Astrophysics Group, Department of Physics \& Astronomy, University of Exeter, Stocker Road, Exeter, EX4 4QL, UK}

\author[0000-0002-3003-3183]{Tyler Gardner}
\affiliation{Astrophysics Group, Department of Physics \& Astronomy, University of Exeter, Stocker Road, Exeter, EX4 4QL, UK}

\author{Mayra Gutierrez}
\affiliation{Department of Astronomy, University of Michigan, Ann Arbor, MI 48109, USA}

\author{Rainer K\"{o}hler}
\affiliation{The CHARA Array of Georgia State University, Mount Wilson Observatory, Mount Wilson, CA 91023, USA}

\author{Karolina Kubiak}
\affiliation{The CHARA Array of Georgia State University, Mount Wilson Observatory, Mount Wilson, CA 91023, USA}

\author[0000-0001-9745-5834]{Cyprien Lanthermann}
\affiliation{The CHARA Array of Georgia State University, Mount Wilson Observatory, Mount Wilson, CA 91023, USA}

\author{Olli Majoinen} 
\affiliation{The CHARA Array of Georgia State University, Mount Wilson Observatory, Mount Wilson, CA 91023, USA}

\author[0000-0003-1038-9702]{Nicholas J. Scott}
\affiliation{The CHARA Array of Georgia State University, Mount Wilson Observatory, Mount Wilson, CA 91023, USA}

\author{Wolfgang Vollmann}
\affiliation{Bundesdeutsche Arbeitsgemeinschaft Veraenderliche Sterne, Munsterdamm 90, D-12169 Berlin, Germany} 
\affiliation{American Association of Variable Star Observers, 185 Alewife Brook Parkway, \#410, Cambridge, MA 02138, USA}

\begin{abstract} 
Stars with initial masses larger than 8 solar masses undergo substantial mass loss through mechanisms that remain elusive. Unraveling the origins of this mass loss is important for comprehending the evolutionary path of these stars,  the type of supernova explosion and  whether they become neutron stars or black hole remnants.
In 2022 December, RW~Cep experienced the Great Dimming in its visible brightness, presenting a unique opportunity to understand mass loss mechanisms. Our previous observations of RW~Cep from the CHARA Array, taken during the dimming phase, show a compelling asymmetry in the star images, with a darker zone on the west side of the star indicating presence of dust in front of the star in our line of sight.  Here, we present multi-epoch observations from CHARA  while the star re-brightened in 2023. We created images using three image reconstruction methods and an analytical model fit.
Comparisons of images acquired during the dimming and re-brightening phases reveal remarkable differences. Specifically, the west side of RW~Cep, initially obscured during the dimming phase, reappeared during the subsequent re-brightening phase and  the measured angular diameter became larger by 8\%. We also observed image changes from epoch to epoch while the star is brightening indicating the time evolution of dust in front of the star.  We suggest that the dimming of RW~Cep was a result from a recent surface mass ejection event, generating a dust cloud that partially obstructed the stellar photosphere.
\end{abstract}


\section{Introduction} \label{sec:intro}

Studying the origins of mass loss rates in massive stars is crucial for understanding their evolutionary trajectory on the Hertzsprung-Russell (H-R) diagram. Mass loss plays a pivotal role in the transformation of these stars into hot supergiants, such as luminous blue variables, B[e] supergiants, or Wolf-Rayet stars. It also determines the type of supernova explosion they will undergo and their subsequent fate as either neutron star or black hole remnants \citep{deJager1998, Langer2012ARA&A..50..107L,Smith2014ARA&A..52..487S}. Recent Great Dimming events observed in Betelgeuse \citep{Montarges2021, Dupree2020, Dupree2022, Humphreys2022} and RW~Cep  \citep[HD~212466,][]{Jones2023, Anugu2023} suggest that episodic surface mass ejections are mainstream processes in supergiants and hypergiants, driven probably by the combined activity of hot convection cells and pulsation. 

\begin{figure*}[h!]
\centering
\includegraphics[width=\textwidth]{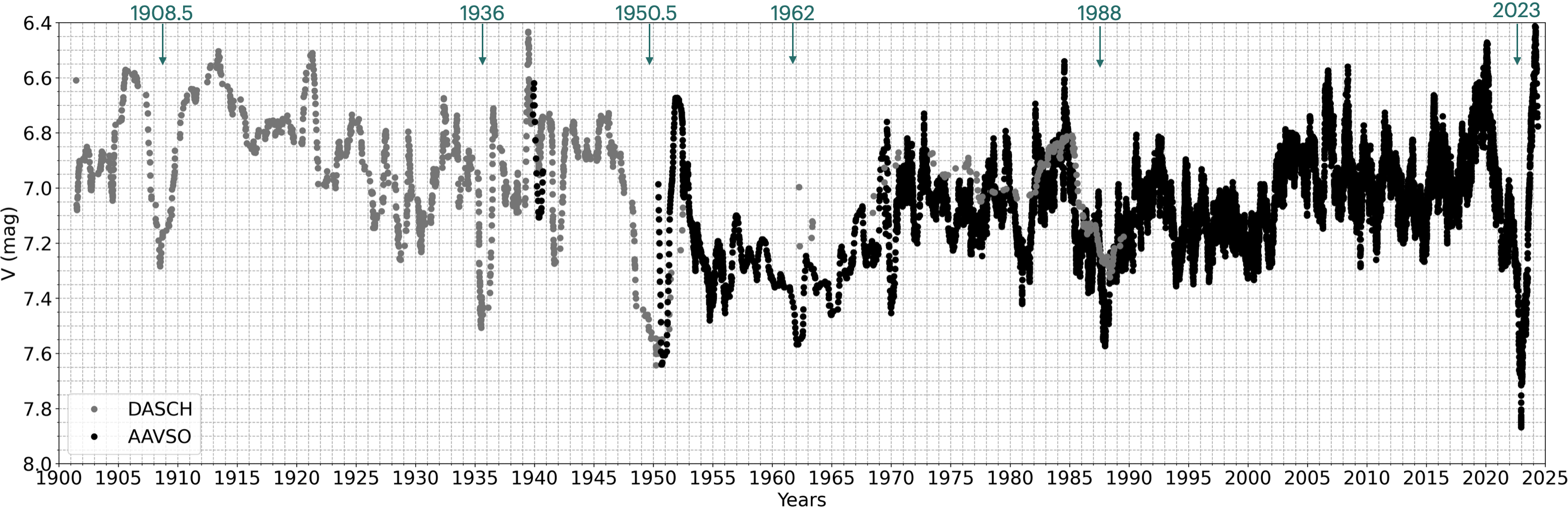}
\caption{The visual brightness of RW~Cep as recorded from 1900 to 2024. Gray data points represent data from DASCH, while black points represent data from AAVSO. To ensure consistency, we adjusted the DASCH data by applying a -2.32 magnitude offset to match the AAVSO data. The dimming event recorded in  2022 December is the faintest observed, compared to all previous dimming events. Another prominent outburst event is evident in 1951, although the data appear noisy. }
\label{Fig:dimming_light_curve_total}
\end{figure*}

\begin{figure*}[h!]
\centering
\includegraphics[width=\textwidth]{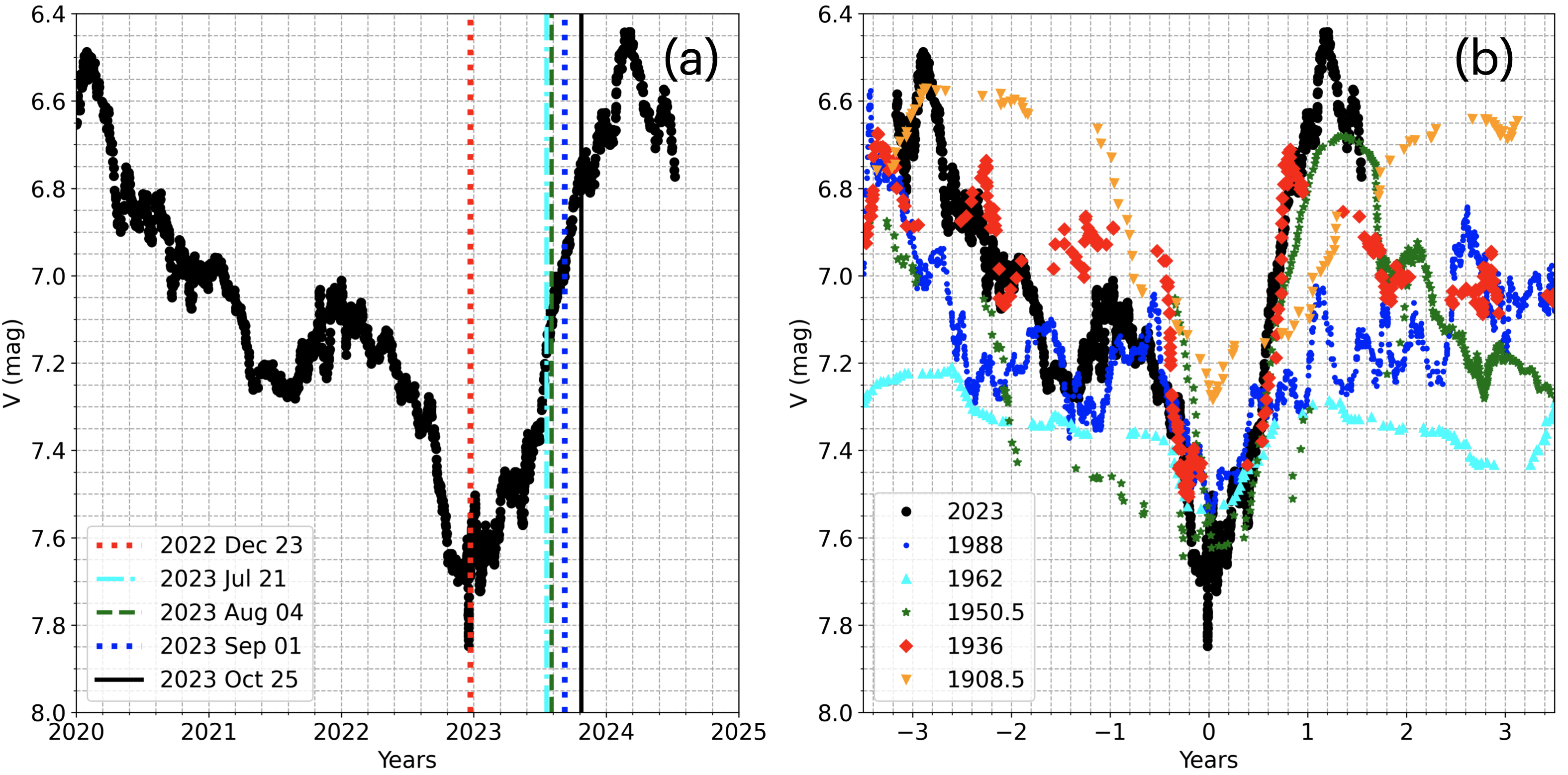}
\caption{
Panel (a): The visual brightness variations of RW~Cep, spanning from January 2020 to May 2024. The vertical lines indicate the specific epochs during which we conducted our CHARA interferometric observations (detailed information on the full light curve can be found in Section~\ref{Sec:aavso_light_curve}). The Great Dimming began at the start of 2020 and reached its faintest level at the end of 2022. Subsequently, it started re-brightening at the beginning of 2023 and returned to its normal brightness by 2024. Interestingly, it began dimming again in March 2024. Panel (b): All the dimming events are centered at the faintest magnitude to compare the fall and rise times of major dimming events since 1900. The dimming event that occurred in 1962 has the longest fall time, while the one in 1908.5 has the shortest fall time. The remaining events have an average fall time of around three years.
}
\label{Fig:dimming_light_curve}
\end{figure*}

This study focuses on RW~Cep, a star that experienced a dramatic dimming event in 2020. As shown in Figures~\ref{Fig:dimming_light_curve_total} and \ref{Fig:dimming_light_curve}, the event manifested as a 1.2-magnitude decrease in brightness  from the beginning of 2020 to the end of 2022. This was followed by a subsequent brightening that began in early 2023 and continued until February 2024. Although RW~Cep experienced six dimming events in the past century, the one recorded in 2022 was the most profound. The star reached its faintest level in over a century, thus earning the name ``The Great Dimming of RW~Cep".  \citet{Jones2023} analyzed the Spectral Energy Distribution (SED) of RW~Cep, suggesting the Great Dimming is part of a series of mass ejections that have likely occurred over the past century.
Analysis of SED reveals the presence of two dust shells: an inner shell with a temperature of 250~K and an outer shell at 100~K. These shells were subsequently imaged by the Multiple Mirror Telescope (MMT) observatory at a wavelength of $11.9 \mu$m, revealing an angular radius of 300-400 mas \citep{Jones2023}. These observations from both SED analysis and direct imaging provide strong evidence that the ongoing fading event is likely part of a series of mass ejection and dust formation episodes in the history of RW~Cep.
\citet{Shenoy2016} also estimated the mass loss rate  $\sim7\times10^{-6}~M_\odot{\rm yr}^{-1}$ in RW~Cep using a DUSTY model fit. Assuming a wind velocity of 50 km~s$^{-1}$ for the dust shells, they further estimated that the shells were likely formed from material ejected during outbursts that occurred $\sim95-140$ years ago. 

High spectral resolution monitoring of the H$\alpha$ spectral line reveals strong emission during the dimming minimum, compared to both before and after the dimming \citep{Kasikov2024}. The H$\alpha$ line is blue-shifted (-100~km/s), indicating that the ejected gas cloud is moving towards us. This gas cloud may be the source of the dust that blocked the star in our line of sight during the dimming.

The analysis of near-infrared (NIR) spectroscopy obtained during the dimming event, as discussed by \citet[][hereafter Paper I]{Anugu2023}, suggests the formation of newly formed dust. This conclusion is drawn from the interpretation of the change in continuum slope, as described in Paper I. The study also found that there was a temperature drop of $\Delta T_{\rm eff} = 300$~K during the dimming event, with an effective temperature $T_{\rm eff}$ of 3900~K compared to 4200~K before the dimming. The interferometric observations of RW~Cep from the same study were conducted by the Georgia State University Center for High Angular Resolution Astronomy (CHARA) Array in both H and K-band wavelengths \citep{tenBrummelaar2005, Schaefer2020}. These observations revealed RW~Cep to be one of the largest stars known, with a limb-darkened diameter of $\theta_{LD} = 2.45$ mas, corresponding to a linear radius of $900 - 1760~R_\odot$ considering its distance range of 3.4 to 6.7 kpc.  Image reconstructions of the CHARA data at H and K-band wavelengths revealed a remarkable asymmetry in the western part of the star, which appeared darker, particularly in the K-band images. This observation hints at a potential association with cool circumstellar dust obscuring the star in our line of sight. See Table~\ref{Table:properties} for a summary of  properties of RW~Cep.

\begin{figure*}[h!]
\centering
\includegraphics[width=\textwidth]{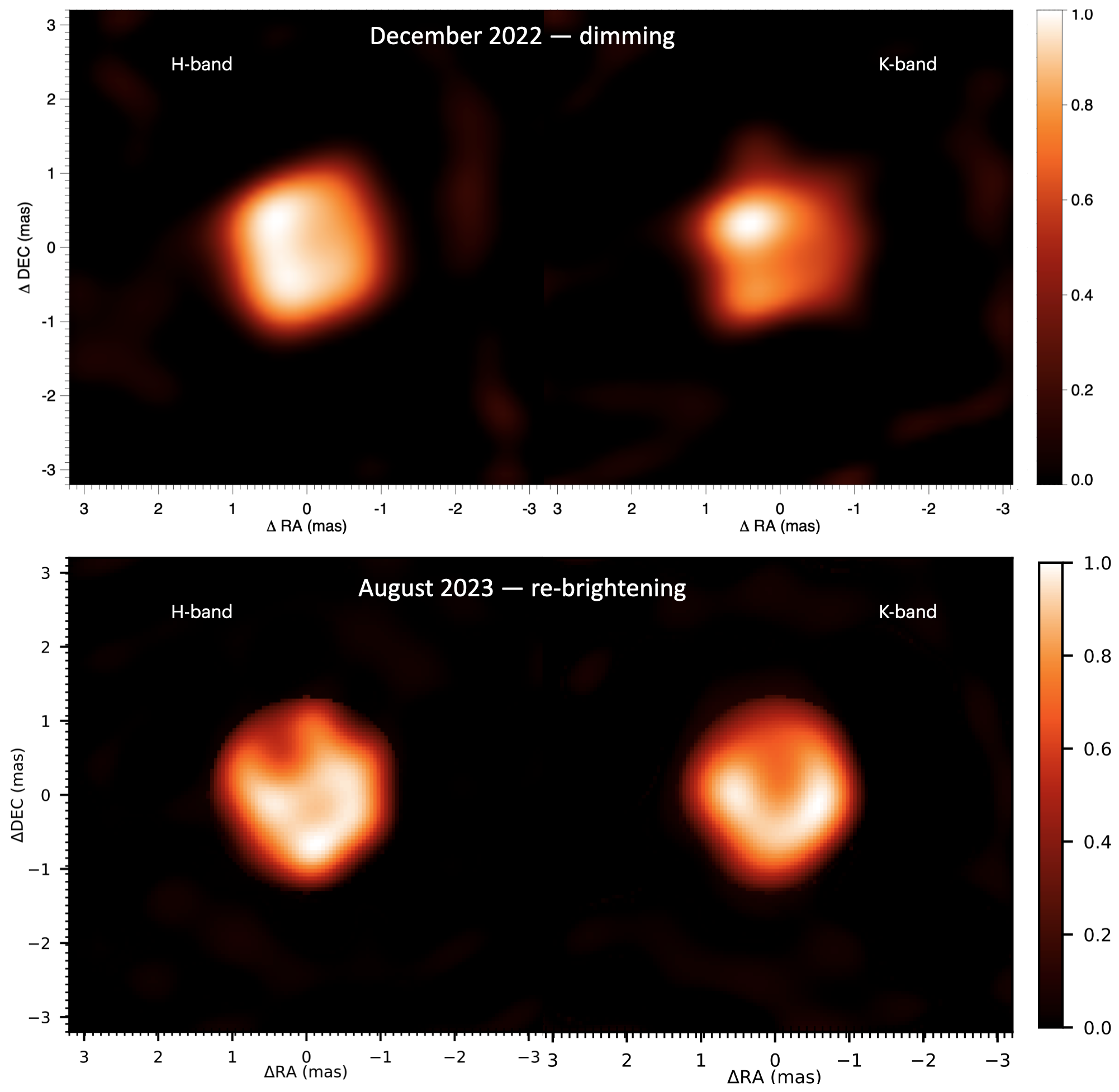}
\caption{Four images of RW~Cep that were obtained from the CHARA Array observations conducted on 2022 December 23 (top panels)  and 2023 August 4 (bottom panels)  in the H and K-band wavelengths. The left panels are from the MIRC-X H-band, while the right panels are from the MYSTIC K-band. The orientation of the images is north pointing upwards and the east towards the left. These images were produced using the image reconstruction techniques described in Section~\ref{Sec:image_recon_conti}. The images are plotted with linear intensity scale.
}
\label{Fig:compare_images_2022_2023}
\end{figure*}

Here we present new CHARA observations that were made while RW~Cep was re-brightening in 2023. During this re-brightening phase, we conducted observations in four epochs spanning almost four months, starting from 2023 July to October.  The images captured during the 2023 re-brightening phase offer a different perspective on RW~Cep, as the western part of the star reappeared bright, standing in stark contrast to the images obtained during the dimming phase (see Figure~\ref{Fig:compare_images_2022_2023}).

This paper is organized into the following sections. The long-term photometric variability of RW~Cep since the beginning of the 20th century is presented in Section~\ref{Sec:aavso_light_curve}. A comparison of the duration of each dimming event (brightness fall and rise time) is also provided in this section. The observation log of our new interferometric observations taken in 2023 using the CHARA Array is described in Section~\ref{Sec:Observations} and Appendix~\ref{Sec:Appendix_log_fit}. This section outlines the data acquisition process and the subsequent data reduction techniques employed to prepare the data for analysis. The initial interpretation of the data using a model fitting technique is presented in Section~\ref{Sec:pmoired_image}. To reconstruct model-independent images, we employ three independent image reconstruction software packages, as described in Sections~\ref{Sec:image_recon_conti}, \ref{Sec:image_CO_lines} and Appendix~\ref{Sec:ROTIR_SURFING}. These algorithms allow us to reconstruct images of RW~Cep, both in the continuum and in specific CO band emission lines. We discuss the results in Section~\ref{Sec:Discussion}, explaining the main processes behind the 2022 dimming. Concluding remarks are presented in the final Section~\ref{Sec:Summary}.

\begin{deluxetable*}{l l l l l l  }
\tablecaption{RW~Cep  properties.}
\label{Table:properties}
\tablewidth{0pt}
\tablehead{
\colhead{Parameter} &  \colhead{Value} &  \colhead{Reference} 
}
\startdata
Stellar Radius (mas) & $1.30\pm0.02$ & This work\\
Stellar Radius ($R_\odot$) & $1100\pm44$ & This work\\
Circumstellar env radius (mas) & $2.15\pm0.10$ & This work \& Paper I \\
Circumstellar env radius ($R_\odot$) & $1820\pm110$ & This work\\
Spectral type & K2Ia-0 - M2Ia-0 & \citet{Keenan1989,Watson2006}\\
Distance (pc) & $3935 \pm 152$  & \citet{Delgado2013, Cantat-Gaudin2020}; \\
& &\citet{Almeida2023, Hunt2024}\\
Temperature (K) & 4200-4400 & before dimming; Paper I, \cite{Jones2023}\\
& 3900  & during dimming, \citet{Anugu2023} \\
Luminosity ($L/L_\odot\times10^3$) & $300$ & \citet{Jones2023} \\
\enddata
\end{deluxetable*}

\section{Historical outburst and dimming activity of RW~Cep}\label{Sec:aavso_light_curve}

Figure~\ref{Fig:dimming_light_curve_total} illustrates 125 years of visual light curve data for RW~Cep covering the period from 1900 to 2024. The plot is generated using archival time-series photometry data obtained from the American Association for Variable Star Observers \citep[AAVSO, ][]{Kloppenborg2023} and the Digital Access to Smithsonian Collections History   \citep[DASCH, ][]{Grindlay2012} databases. We applied a -2.32 magnitude offset to the DASCH dataset to match the AAVSO dataset. To filter the noisy observations, a 15-day moving average was applied to smooth the data using a low-pass filter \textit{signal.lfilter} from Python, and subsequently, the data were binned into 7-day intervals.

\begin{figure*}[h]
\centering
\includegraphics[width=\textwidth]{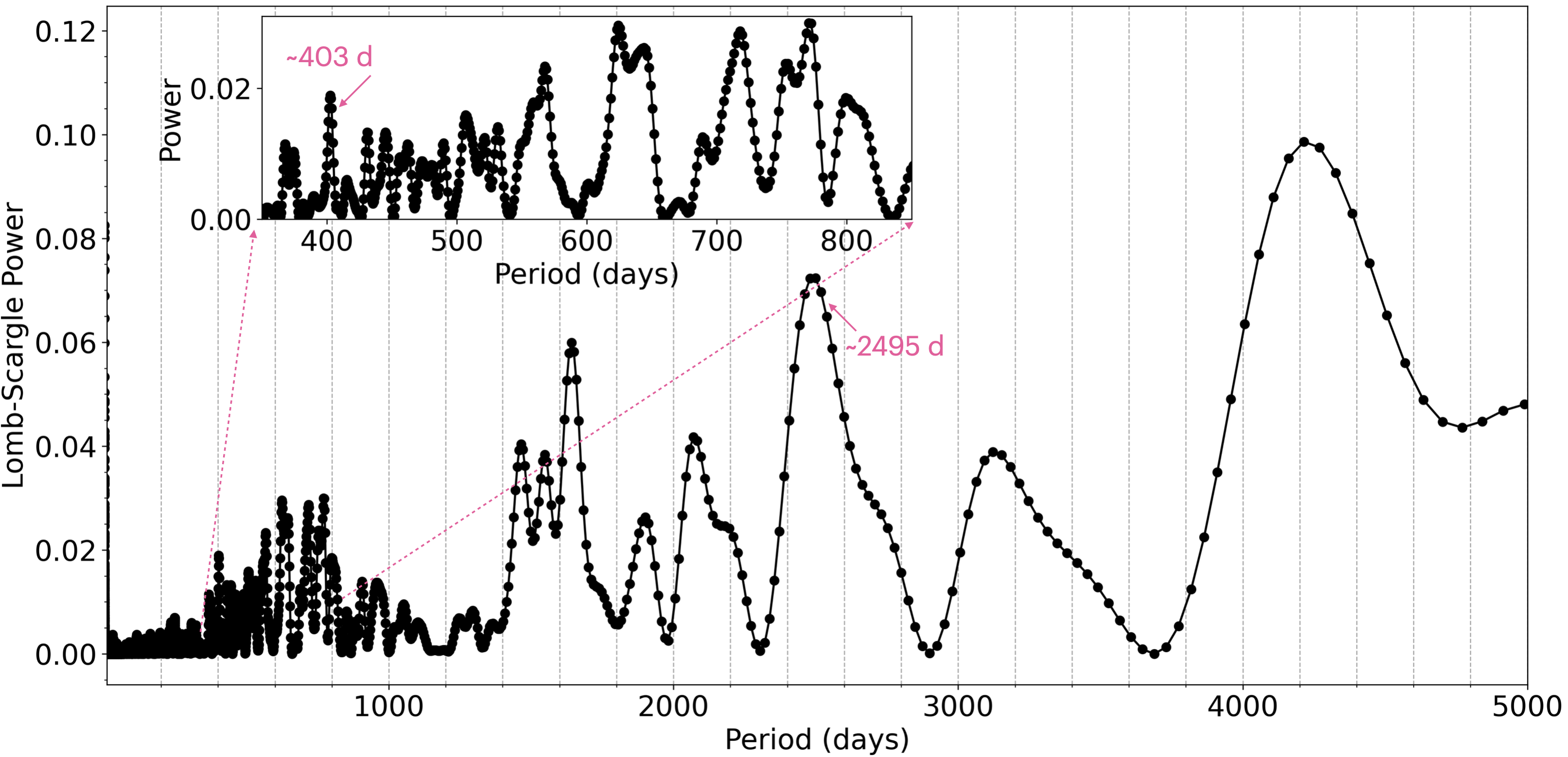}
\caption{The Lomb-Scargle periodogram was computed from the AAVSO data. The inset image details the region near the pulsation peak. The pulsation peak occurs at $403\pm20$ days, and the long secondary period is $2495\pm30$ days, equivalent to 6.8 years.}
\label{Fig:dimming_light_curve_period}
\end{figure*}

A visual examination of the light curve shows a semi-episodic pattern of dimming and re-brightening activity. We identify six significant dimming episodes since 1900, occurring in 1908.5, 1936, 1950.5 \citep{Semakin1954PZ.....10..191S}, 1962, 1988, and most recently in 2023. These events were identified based on two criteria: (i) the dimming event must cause a significant decrease in brightness, exceeding the pulsational instability dimming of $\pm0.3$ magnitudes, and (ii) the dimming event should have a longer fall and rise time, meaning the time between the start of dimming and the subsequent return to normal brightness must be longer than the $\sim1-1.5$ year pulsation period of the star (see Figure~\ref{Fig:dimming_light_curve_period}). In these identified events, the visual brightness dropped by up to $\Delta V=1.2$ magnitudes, and the fall and rise times of the dimming events were over four years (see Figure~\ref{Fig:dimming_light_curve}b). RW Cep exhibits asymmetric dimming light curves similar to [W60]B90 in Large Magellanic Cloud \citep{Munoz-Sanchez2024arXiv240511019M}. This means the star takes longer time to decrease in brightness (fall) than it does to increase in brightness (rise).

We propose that these major dimming events resulted from outbursts, leading to substantial mass loss with the release of gas clouds from the stellar surface. The analysis of SED and thermal imaging from the MMT reveals that presence of two dust shells, which could have formed from these mass ejected material \citep{Jones2023}. As these gas clouds travel outward, they cool down and condense into dust, temporarily blocking the star to dim in our line of sight before eventually dispersing. This formation of dust would naturally lead to a decrease in brightness, with the brightness eventually returning to normal levels once the dust cloud has moved away. Interestingly, the 2020-2022 event was the faintest ($\Delta V=1.2$, Figure~\ref{Fig:dimming_light_curve}b) compared to all other events, suggesting the name ``The Great Dimming." The next major event occurred around mid-1950, although the photometric observations appear noisy. The shortest period between two dimming events ever recorded is around 14.5 years, which occurred between 1936 and 1950.5. The longest period between two dimming is around 35 years, stretching from 1988 to 2023. On the other hand, the small duration (fall and rise time) dimming events may have been triggered by pulsational instabilities within the extensive, tenuous outer envelope of the star \citep{Stothers1969ApJ...156..541S, vanGenderen2019}.

Figure \ref{Fig:dimming_light_curve_period} presents the power spectrum periodogram of RW~Cep. We utilized the Lomb-Scargle periodogram module from the Python astropy.timeseries library \citep{VanderPlas2018ApJS..236...16V} to compute the periodogram based on the AAVSO data. The Lomb-Scargle periodogram is a statistical tool designed to detect periodic signals in unevenly spaced data. In the power spectrum, several peaks were observed, ranging from a few hundred days to more than 10 years. Theoretical predictions suggest that multiple periodicities may arise from multimode pulsations, as observed in red supergiant stars and yellow hypergiants \citep{Kiss2006MNRAS.372.1721K,vanGenderen2019}. We carefully examined all the peaks in the power spectrum to determine if they were harmonics of long secondary periods or aliasing peaks resulting from observing window gaps in the data. RW~Cep exhibits no unique pulsation period, but varies between 350 and 800 days. The frequently observed pulsation period is $403\pm20$ days, while the long secondary period spans approximately $2495\pm30$ days.

The long secondary periods maybe linked to the turnover duration time of large convection cells on the stellar surface, which are recognized for their substantial depth and vigorous convection \citep{Stothers1971A&A....10..290S,Stothers2010ApJ...725.1170S,Stothers2012}. Observing RW~Cep with the CHARA Array throughout this long secondary period could serve to corroborate the relationship between the convection cells and the long second period variability.

\section{The CHARA Array observations in H and K-band wavelengths}\label{Sec:Observations}

In this section, we briefly describe the 2023 CHARA observing campaign, the data reduction and the data calibration. A detailed observing log is provided in Appendix~\ref{Sec:Appendix_log_fit}. We conducted a four-epoch observational campaign, with each epoch separated by approximately one month. Figure~\ref{Fig:dimming_light_curve} marks the corresponding CHARA observing dates on the visual light curve. We followed  similar observing and data reduction procedures as outlined in Paper I. The observations were carried out using the CHARA Array located at the Mount Wilson, California, USA. The CHARA Array is composed of six 1-meter telescopes arranged in a Y-shaped configuration \citep{tenBrummelaar2005,Schaefer2020}. Each telescope is equipped with an adaptive optics system to improve the flux injection into the beam combiner instruments in the laboratory \citep{Che2013JAI.....240007C,Anugu2020SPIE11446E..22A}. The CHARA Array stands as the largest optical and near-infrared interferometer in the world, featuring baseline lengths ranging from $B=34$~m to 331~m. We used the H-band MIRC-X \citep[$\lambda=1.5 - 1.8 ~\mu$m,][]{Anugu2020} and the K-band MYSTIC instruments \citep[$\lambda=2.0 - 2.39 ~\mu$m,][]{Setterholm2023} to collect interferometric data. MIRC-X and MYSTIC simultaneously combine six telescope beams, providing 15 squared visibilities ($V^2$) and 10 independent closure phases ($T3PHI$) within each operational wavelength band. The MIRC-X ($\mathcal{R}=190$) and MYSTIC ($\mathcal{R}=278$) observations cover 30 and 52 spectral channels in H and K-bands and they deliver angular resolutions of approximately $\lambda/2B \sim 0.5$ and 0.6 milliarcseconds (mas), respectively.

The raw MIRC-X and MYSTIC data sets\footnote{\href{https://www.chara.gsu.edu/observers/database}{can be accessed from https://www.chara.gsu.edu/observers/database}}  were reduced using the standard MIRC-X~pipeline, version 1.4.0 \citep{Anugu2020,le_bouquin_2024_12735292}.
The pipeline produces squared visibilities ($V^2$) and closure phases ($T3PHI$) and flux.
We applied the IDL mircx\_cal.script routine to calibrate the target data with calibrator data, ensuring accuracy and reliability in our analysis.

\begin{figure*}[h!]
\centering
\includegraphics[width=\textwidth]{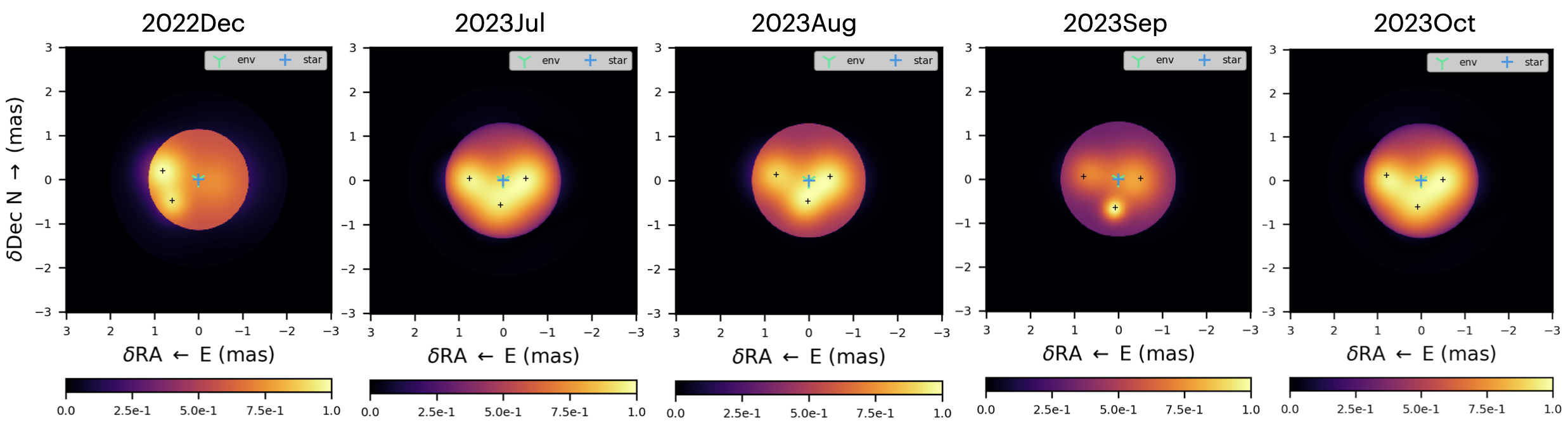}
\caption{The PMOIRED best-fit model fitting of K-band datasets for the epochs starting from 2022 December to 2023 October. The images are centered at $\lambda \sim 2.2~\mu$m. The image colors are plotted on a linear intensity scale. The colored ``+” and ``Y” symbols indicate the center of the star and the circumstellar envelope, while the black ``+” marks denote the locations of the bright spots. }
\label{Fig:pmoired_images_all_2023_epochs}
\end{figure*}

\begin{figure}[h!]
\centering
\includegraphics[width=0.49\textwidth]{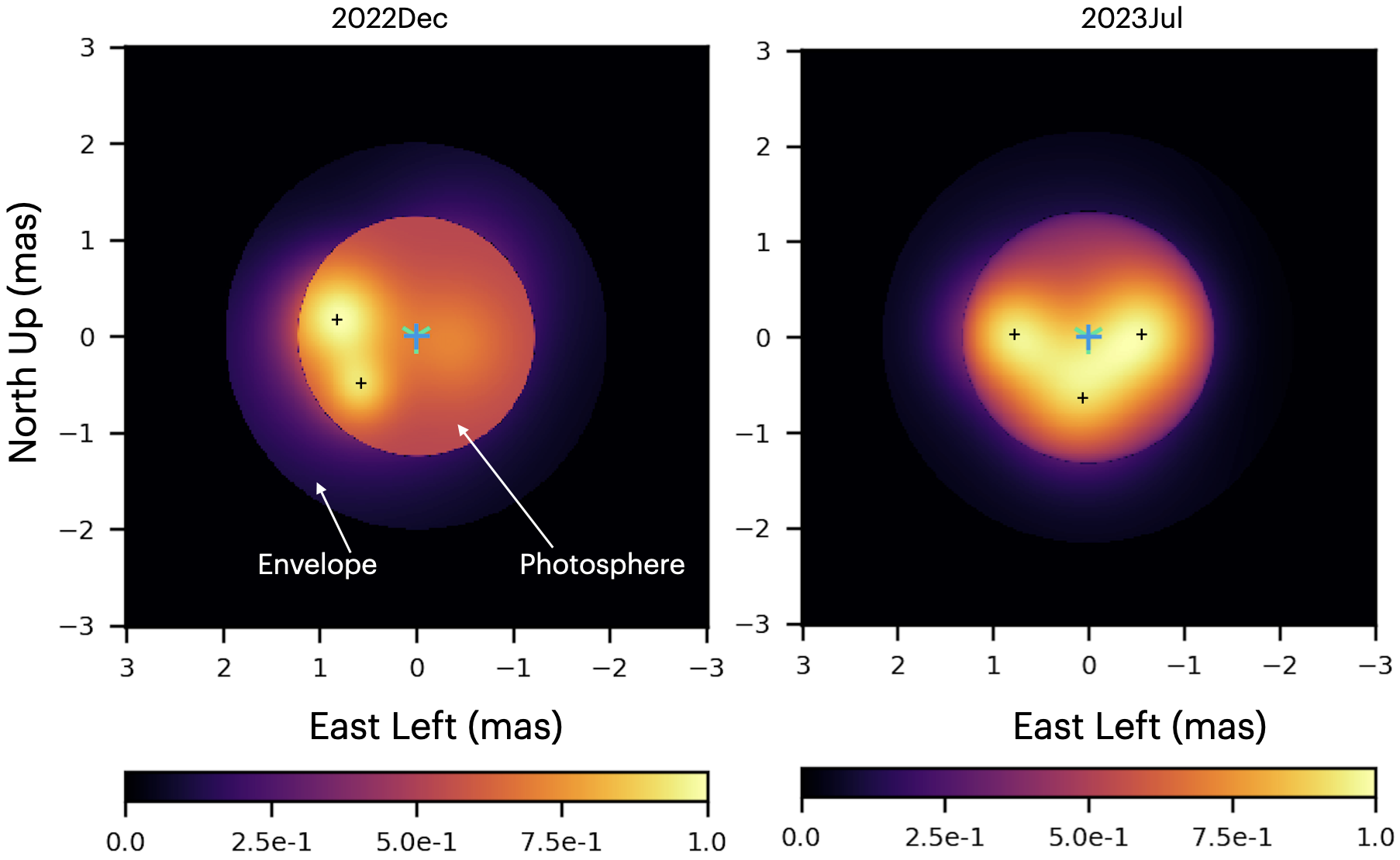}
\caption{The best-fit model fitting of K-band at $\lambda=2.35~\mu$m using the PMOIRED software is presented for the epochs during dimming on 2022 December 23 (left panel) and re-brightening on 2023 July 21 (right panel). The image colors are plotted on a linear intensity scale. The symbols on the image represent the same elements as in Figure~\ref{Fig:pmoired_images_all_2023_epochs}.
}
\label{Fig:pmoired_images_compare_2022_2023}
\end{figure}

\begin{figure*}[h!]
\centering
\includegraphics[width=\textwidth]{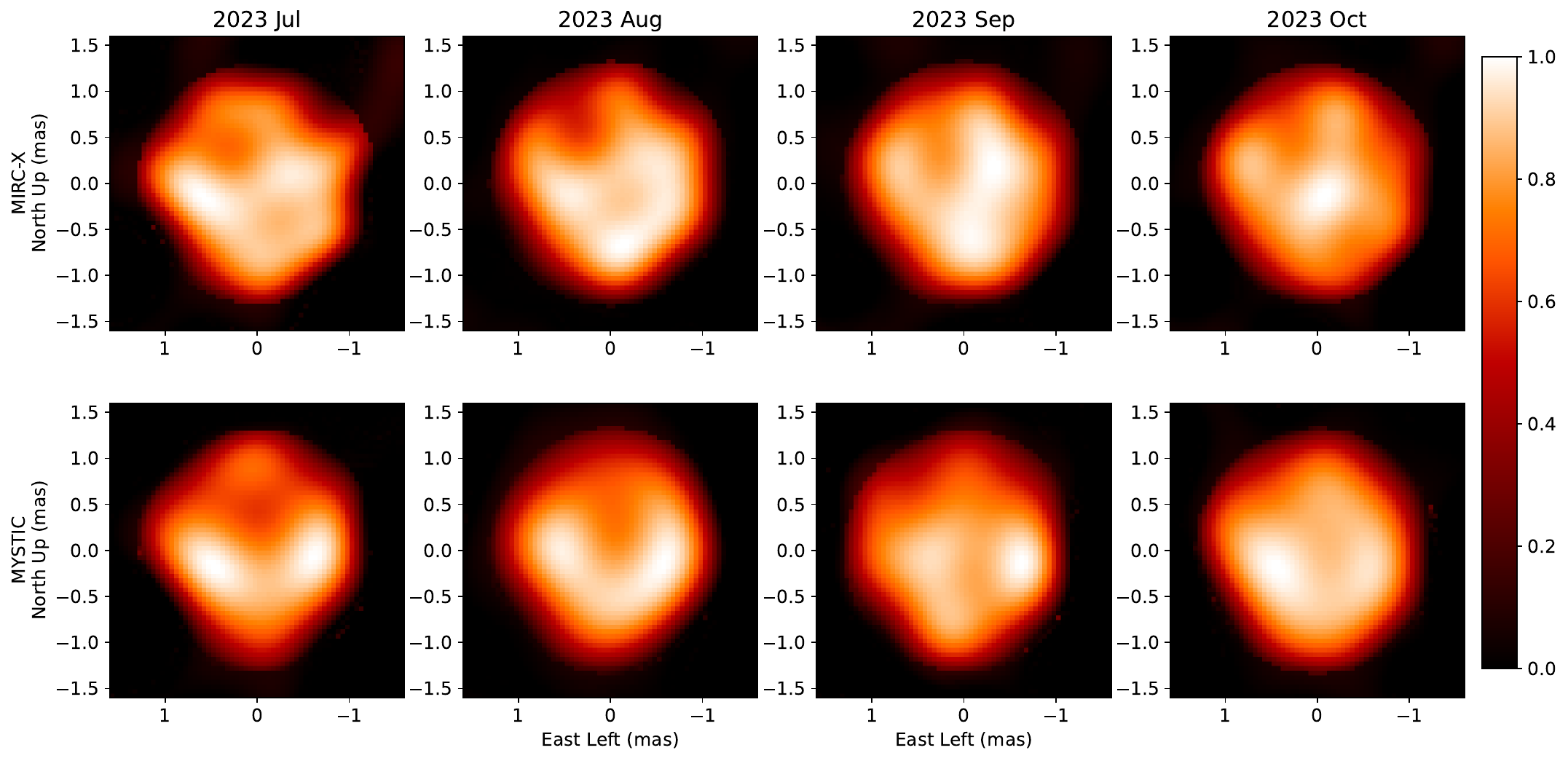}
\caption{The OITOOLS reconstructed images of RW~Cep for epochs 2023 July, August, September and October while the star is re-brightened. The top panel corresponds to the H-band MIRC-X, while the bottom panel corresponds to the K-band MYSTIC.  Pixel scale is 0.05 mas. The linear intensity scale is employed for plotting all images. The stellar surface appears asymmetric, deviating from a circular shape. The K-band images show three bright spots arranged in a V-shape on the star surface comparable to the PMOIRED images (Figure~\ref{Fig:pmoired_images_all_2023_epochs}), that we interpret as giant convection cells. The H-band images, with ($\times1.4$) higher resolution, reveal more structures than the K-band images. The darker regions on the star surface correspond to cool, infalling material located between these giant convection cells. Additionally, small variations are observed in the images from epoch to epoch; these changes could be attributed to the time-varying dispersal of dust away from the star.} 
\label{Fig:oitools_images}
\end{figure*}

\begin{figure*}[h]
\centering
\includegraphics[width=\textwidth]{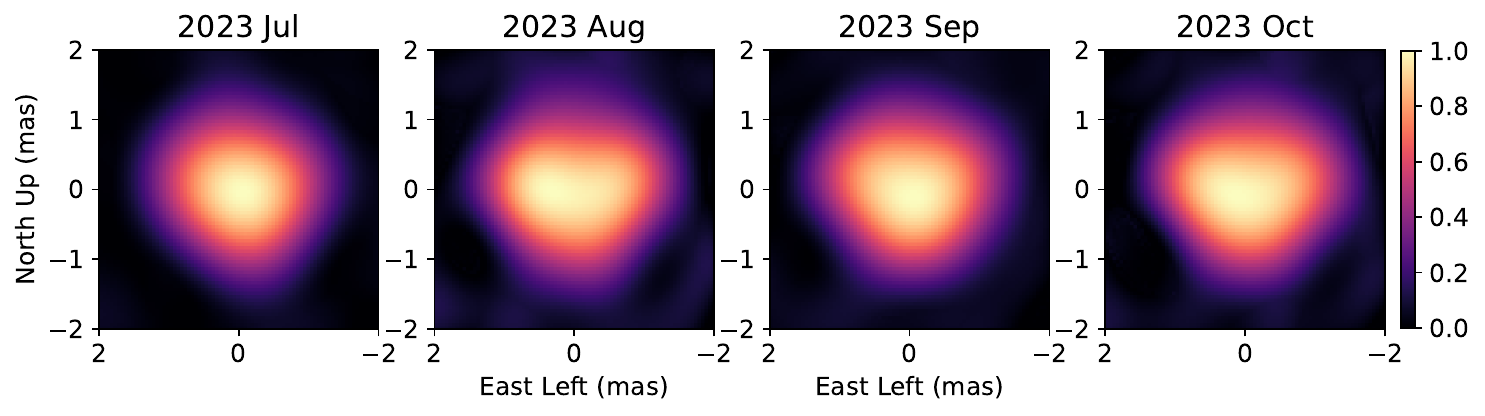}
\caption{OITOOLS reconstructed images of RW~Cep in the CO bands ($\lambda=2.31-2.39\mu$m). Each image corresponds to a distinct epoch, which is indicated at the top of each panel. The intensities in these images are depicted on a linear intensity scale. In these reconstructions, the central bright region extending 2.6~mas corresponds to the stellar photosphere, while the outer region encapsulates the circumstellar envelope characterized by CO bands. 
}
\label{Fig:CO_lines_images}
\end{figure*}

\section{Model Fitting of the Continuum and Line Regions}\label{Sec:pmoired_image}

To interpret the $V^2$ and $T3PHI$ data, we applied model fitting  utilizing the Python-based Parametric Modeling of Optical InteRferomEtric Data software, PMOIRED~\citep{Merand2022,antoine_merand_2024_10889235}. We used MYSTIC data for this model fitting to model bright spots on the surface of the star and CO spectral lines. The PMOIRED software enables the construction of a complex model in a linear combination of simple components. Additionally, it is equipped with grid search mechanisms to explore parameter space, to check for local minima traps, and facilitate finding the global minimum.

For this model fitting, we used MYSTIC K-band data including all wavelengths $\lambda=2.00-2.39~\mu$m. Our fitting procedure began by establishing parameters for the limb-darkened disk diameter and an extended circumstellar envelope surrounding the star. To model structures on the stellar surface, initially, a single bright circular Gaussian spot was introduced. Watching the reduced chi-squared $\chi_{\rm red}^2$, we implemented multiple bright spots denoted by $n$. Each spot was characterized using a Gaussian model defined by four free parameters: Full Width at Half Maximum (FWHM, $d_n$), the  relative flux of spot ($f_n$) in comparison to the star, and its position (projected radial distance $r_n$, position angle $\theta_n$) on the stellar surface relative to the center of the star. Other parameters included the limb-darkening disk diameter ($d_{\rm photo}$), limb-darkening power law coefficient ($\alpha$), and flux ($f_{\star}$) of the central star. The overall flux of the system remained fixed at 1.0, while the circumstellar environment flux ($f_{\rm env}$) was computed as $1-f_\star  - \sum_{n=1}^{n=3}{f_n}$. Although the circumstellar envelope is spherical in shape, we adapted a simple 2-dimensional projection disk characterized by inner and outer diameter and radial profile parameters. The  outer diameter of the circumstellar envelope ($d_{\rm env}$) is searched for within the range between the photosphere diameter ($d_{\rm photo}$) and $2 \times d_{\rm photo}$. The inner diameter of the circumstellar envelope is set to match the size of the star ($d_{\rm photo}$). We also fitted the CO spectral lines. We used the near-infrared spectrum from the APO observatory, as described in Paper I, to perform a simultaneous fit. We used the known CO band-head wavelengths to set the boundary between the continuum and CO line regions, and we adopted a truncated,
exponential spatial profile in PMOIRED to represent the extended atmosphere. For simplicity, we assumed coplanarity between the star and circumstellar envelope. 

In determining the positions of these spots, we used the grid search method from the PMOIRED software, adhering to the following constraints. Firstly, the location of a spot $d_n$ must be within the boundary of the stellar diameter, ensuring that $d_n < d_{\rm photo}/2$. Secondly, the diameter of each spot must not be larger than the diameter of the star. We also fitted the spectral lines, revealing circumstellar envelope absorption spectra with approximately 35\% absorption.

Through this modeling approach, we identified three distinct bright spots $n=3$, achieving a $\chi_{\rm red}^2 = 3-7$ for different epochs (See Table~\ref{Table:fit_chi2}). Attempts to add a fourth bright spot did not significantly improve $\chi_{\rm red}^2$ value.  The final model yielded a limb-darkened star, with three Gaussian spots and an extended circumstellar 
 envelope for the wavelength range of the CO lines 
($\lambda > 2.31 \mu$m).

Figure~\ref{Fig:pmoired_images_all_2023_epochs} shows the best-fit analytical model images for all imaging epochs covering both dimming and re-brightening phases. In this paper all the images are plotted on a linear intensity scale.  From these images, it is evident that during the dimming phase (2022 December), the western part of the star appeared darker compared to the images obtained in 2023.  Table~\ref{Table:pmoired_param} present the best-fit parameters from the PMOIRED analytical model. In the 2022 December image, two spots (one giant and one smaller) were detected, positioned in the western part of the star. In the 2023 epochs, three giant spots were identified, located in the southern and eastern parts in addition to the western part. These three spots have FWHM sizes of 0.93 mas,  0.92 mas  and 0.72 mas located in the western, eastern and southern parts, respectively. These bright spots contribute 9\%, 8\%, and 4\% of the total flux, respectively. We measured the limb-darkened diameter to be $2.60\pm0.02$ mas with a limb-darkening coefficient of $0.45\pm0.06$.  The background flux is approximately zero within measurement errors.

Figure~\ref{Fig:pmoired_images_compare_2022_2023} presents the circumstellar envelope detection from the longer CO band wavelengths. We found that the circumstellar envelope has a diameter of approximately $4.3\pm0.1$ mas, i.e., the circumstellar envelope extends up to twice the size of the star, which is typical for the MOLsphere of evolved stars \citep[e.g.,][]{Perrin2005, Ohnaka2011, Montarges2014, Anugu2023, Anugu2024}. 

Table~\ref{Table:diameter_fit} shows limb-darkened diameters fitted using three different software packages. The PMOIRED model fit gives a limb-darkened diameter of $2.62 \pm 0.03$ mas. Interestingly, the diameter measured in the 2023 epochs is larger than the diameter measured in  2022 December. Appendix~\ref{Sec:Appendix_log_fit}  shows the K-band ($2.0-2.39~\mu$m) fitting residuals from the PMOIRED model fitting for the epoch of 2023 July 21.

\begin{deluxetable}{c  c c c  c c c  }
\tablecaption{The best-fit $\chi^2_{\rm red}$ of the K-band measurements for each method.}
\label{Table:fit_chi2}
\tablewidth{0pt}
\tablehead{
Epoch &  OITOOLS  &   ROTIR & SURFING & PMOIRED 
}
\startdata
2022 Dec  & 1.26 & 1.37  & 1.32 & 4 \\
2023 Jul  & 2.5 & 2.7  & 1.22 & 5 \\
2023 Aug  & 1.9 & 2.0  & 0.47 & 5 \\
2023 Sep  & 0.8 & 0.9  & 0.68 & 6\\
2023 Oct  & 0.9 & 1.0  & 0.19 & 3
\enddata
\end{deluxetable}

\begin{deluxetable}{l l l l l l l l l }
\tablecaption{The best-fit PMOIRED model fitting parameters from epoch 2023 July}
\label{Table:pmoired_param}
\tablewidth{0pt}
\tablehead{
\colhead{Param} & \colhead{Value}  
}
\startdata
Fluxes & \\
$f_1$ (\%)&             $8\pm 1 $\\
$f_2$ (\%)&             $4\pm 1 $\\
$f_3$ (\%)&             $9\pm 1 $\\
$f_{\rm env}$ (\%)&     $35\pm 3 $\\
\hline 
FWHM of spots & \\
$d_1$ (mas)&           $0.92\pm 0.01 $\\
$d_2$ (mas)&           $0.72\pm 0.04 $\\
$d_3$ (mas)&         $0.93\pm 0.02$ \\
\hline 
Spot position from origin & ($r_n$, $\theta_n$)\\
$r_1$ (mas)&           $0.83\pm 0.01 $ \\
$r_2$ (mas)&           $0.70\pm 0.01 $\\
$r_3$ (mas)&           $0.45\pm 0.02 $\\
$\theta_1$ ($^\circ$)&             $84\pm 1$\\
$\theta_2$ ($^\circ$)&             $167\pm 1 $\\
$\theta_3$($^\circ$)&             $272\pm 1 $\\
\hline 
Limb-darkened angular diameter &  \\
$d_{\rm photo}$ (mas)&             $2.62\pm 0.03 $\\
$\alpha$  &  $0.45 \pm 0.06$ \\
$d_{\rm env}$ (mas)&           $4.36\pm 0.13 $\\
\enddata
\end{deluxetable}

\begin{deluxetable*}{c | c c c | c c c  }
\tablecaption{The limb darkened diameter, $d_{\rm photo}$ (mas), measured for all epochs. }
\label{Table:diameter_fit}
\tablewidth{0pt}
\tablehead{
Epoch  &  & H-band &   &  & K ($2.0-2.3~\mu$m) &  \\
UT Date &  OITOOLS  &  SURFING & PMOIRED & OITOOLS &  SURFING & PMOIRED 
}
\startdata
2022 Dec  & $2.41\pm 0.08$ & 2.44 & $2.45 \pm 0.04$ & $2.35\pm0.13$ & $2.44$ & $2.41 \pm 0.07$\\
2023 Jul  & $2.68\pm0.09$ & 2.72  & $2.62\pm0.05$ & $2.62\pm0.09$ & $2.50$ & $2.62\pm 0.03$\\
2023  Aug & $2.61\pm0.10$ & 2.49  & $2.59\pm 0.05$ & $2.63\pm0.11$ & $2.52$ & $ 2.60\pm0.01$\\
2023 Sep  & $2.62\pm0.11$ & 2.62  & $2.63\pm0.16$ & $2.57\pm0.16$ & $2.58$ & $2.64\pm0.02$\\
2023 Oct  & $2.60\pm0.09$ & 2.59  & $2.59\pm 0.05$ & $2.59\pm0.11$ & $2.57$ & $2.58\pm 0.01$\\
\enddata
\tablecomments{The 2022 December epoch marked the dimming of the star, while the remaining epochs were observed during its re-brightening phase.} 
\end{deluxetable*}





\section{IMAGE RECONSTRUCTION IN CONTINUUM REGIONS}\label{Sec:image_recon_conti}

The image reconstructions are performed separately for the continuum and CO band wavelength datasets. This section elaborates on the continuum in the H-band and K-band ($\lambda=2.0-2.31~\mu$m) wavelength image reconstructions, while Section~\ref{Sec:image_CO_lines} delves into the CO band and continuum ($\lambda = 2.31-2.39~\mu$m) image reconstructions.

We employed similar image reconstruction methods as those employed in Paper I to generate aperture synthesis images. The first code we used was Optical Interferometry TOOLS (OITOOLS) \citep{Norris2021}\footnote{\href{https://github.com/fabienbaron/OITOOLS.jl.git}{https://github.com/fabienbaron/OITOOLS.jl.git}} to compare with the analytical images made with PMOIRED.  OITOOLS employs a quasi-Newton method with rapid gradient minimization using Variable Metric Limited Memory with Bounds (VMLMB). OITOOLS is designed for both two-dimensional image reconstruction and model fitting applications.  This software starts by modeling the best-fit limb-darkened angular diameter, which sets the boundaries for the assigned flux.  OITOOLS is coded in the Julia programming language.

We used the same image reconstruction parameters as in Paper I, including positivity constraints and total variation regularization, with the hyperparameter weight chosen using the classical $L$-curve method. In OITOOLS, we applied edge-preserving $\ell_2-\ell_1$ regularization. The images were constructed with a field of view of $6.4 \times 6.4$ mas and a pixel scale of 0.05 mas. To address the issue of local minima traps, 500 image reconstructions were initiated from randomly generated images. To further assess the robustness of our reconstructions, images were created using different exposure times, such as short 30-second exposures and 10-minute averaged integrations. The consistency and reliability of the resulting images were verified (Figure~{\ref{Fig:oitools_images}).

We also used two other image reconstruction methods, ROTIR \citep{Martinez2021} and SURFING \citep{Roettenbacher2016} to reconstruct images from the interferometric data as presented in Appendix~\ref{Sec:ROTIR_SURFING}. Table~\ref{Table:fit_chi2} presents the $\chi^2_{\rm red}$ values for each epoch using different modeling and image reconstruction methods. The $\chi^2_{\rm red}$ values for PMOIRED are comparatively larger, which may be due to the smaller number of parameters considered compared to the image reconstruction methods.

To ensure the validity of the images and distinguish genuine features from artifacts caused by gaps in the ($u,v$) coverage, we performed additional checks. Furthermore, to explore the potential influence of different imaging and optimization choices within the software, we generated synthetic datasets for further analysis. The OIFITS datasets, mimicking actual observations including $(u,v)$ coverage and signal-to-noise ratios, were used to generate image reconstructions with OITOOLS. The image reconstructions based on the synthetic datasets are displayed in Appendix~\ref{Sec:Synthetic}. These images show no substantial features, indicating that the observed features were not artifacts from the gaps in the $(u,v)$ coverage.

\section{IMAGE RECONSTRUCTION IN THE CO SPECTRAL LINES}\label{Sec:image_CO_lines}

The spectral channels around the CO bandhead  lines ($\lambda=2.31-2.39~\mu$m) were separated from the K-band datasets and we conducted image reconstructions on them using OITOOLS, following the methodologies detailed in Section~\ref{Sec:image_recon_conti}. Figure~\ref{Fig:CO_lines_images} displays the resulting images for different observing epochs, showcasing two distinct components: a central star photosphere and an outer circumstellar envelope with diameter of $4.3\pm0.2$ mas, which closely agree with the model fitting results (see Section~\ref{Sec:pmoired_image}). These images  highlight the asymmetric appearance of the stellar photosphere and circumstellar envelope. 

\section{Comparison of Images Derived from Various Image Reconstruction Techniques}\label{Sec:compare_images}

Upon visually examining the continuum wavelength images produced by the  PMOIRED (Figure~\ref{Fig:pmoired_images_all_2023_epochs}) and OITOOLS (Figure~\ref{Fig:oitools_images}) software together with those from ROTIR (Figure~\ref{Fig:ROTIR_images}) and SURFING (Figure~\ref{Fig:SURFING_images}), we observe a remarkable agreement among the images, with some noteworthy differences. One striking common feature is the asymmetric shape of the star, where the Southern and Northern regions appear much smaller than the equator (see Figure~\ref{Fig:oitools_images}). Another notable pattern on the stellar surface is the orientation of bright features forming in a V-shape. This common pattern across different reconstruction techniques highlights the consistency of the image reconstructions and model fitting.  We also observe some stellar surface image differences between the 2023 July/August epochs in a comparison to the September/October epochs, potentially attributed to physical processes (see discussion in Section~\ref{Sec:Discussion}).



Furthermore, apparent disparities emerge between the images captured in the H and K-band wavelength bands (Figure~\ref{Fig:oitools_images}). These differences primarily result from variations in the effective angular resolutions of the telescope for each band. H-band observations exhibit higher resolution ($\lambda/2B\sim0.5$ mas) compared to the K-band ($\sim0.6$ mas), leading to images with enhanced contrast and detail across the stellar disk.

Table~\ref{Table:diameter_fit} summarizes the limb-darkened diameters measured for each observing epoch using three different software packages.  We used the OITOOLS analytical fit to measure the limb-darkened diameter, as described in Paper I. Additionally, we estimated the limb-darkened diameter using the SURFING algorithm, also detailed in Paper I. The resulting limb-darkened diameters during the dimming and re-brightening phases are presented in Table~\ref{Table:diameter_fit}. During the dimming phase, the average limb-darkened diameter was $2.42\pm0.02$ mas. However, during the re-brightening phase, the average limb-darkened diameter increased to $2.60\pm 0.01$ mas. The stellar diameter increased by $8\pm1$\%  after the dimming event, compared to the dimming event itself. (Paper I). This substantial increase suggests that the stellar surface was being partially blocked during the dimming phase, possibly by circumstellar dust.

\begin{figure}[h]
\centering
\includegraphics[width=0.49\textwidth]{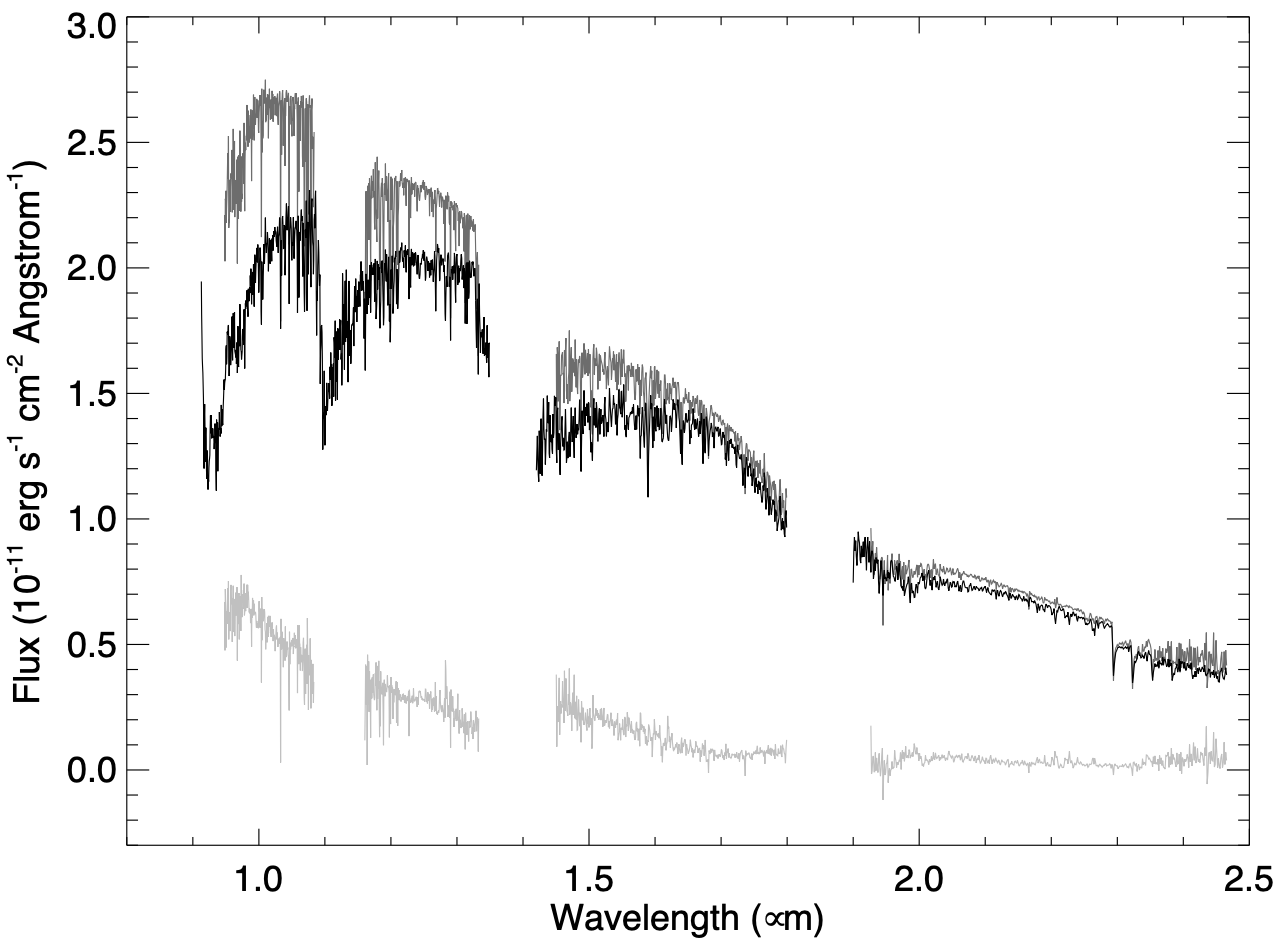}
\caption{The near-infrared spectrum of RW Cep made with the Apache Point Observatory
TripleSpec spectrograph. The black line shows the spectrum obtained close to minimum
brightness during the Great Dimming (observed on 2023 Jan 9 and 12), and
the dark gray line shows the spectrum after the star had almost returned to normal brightness
(observed on 2023 Sep 26).  The difference (bright minus faint state) is shown
in light gray at bottom, and this demonstrates that the Great Dimming had a larger
amplitude at lower wavelengths (as expected for dust extinction).}
\label{Fig:nir_spectrum}
\end{figure}

\begin{figure}[h]
\centering
\includegraphics[width=0.49\textwidth]{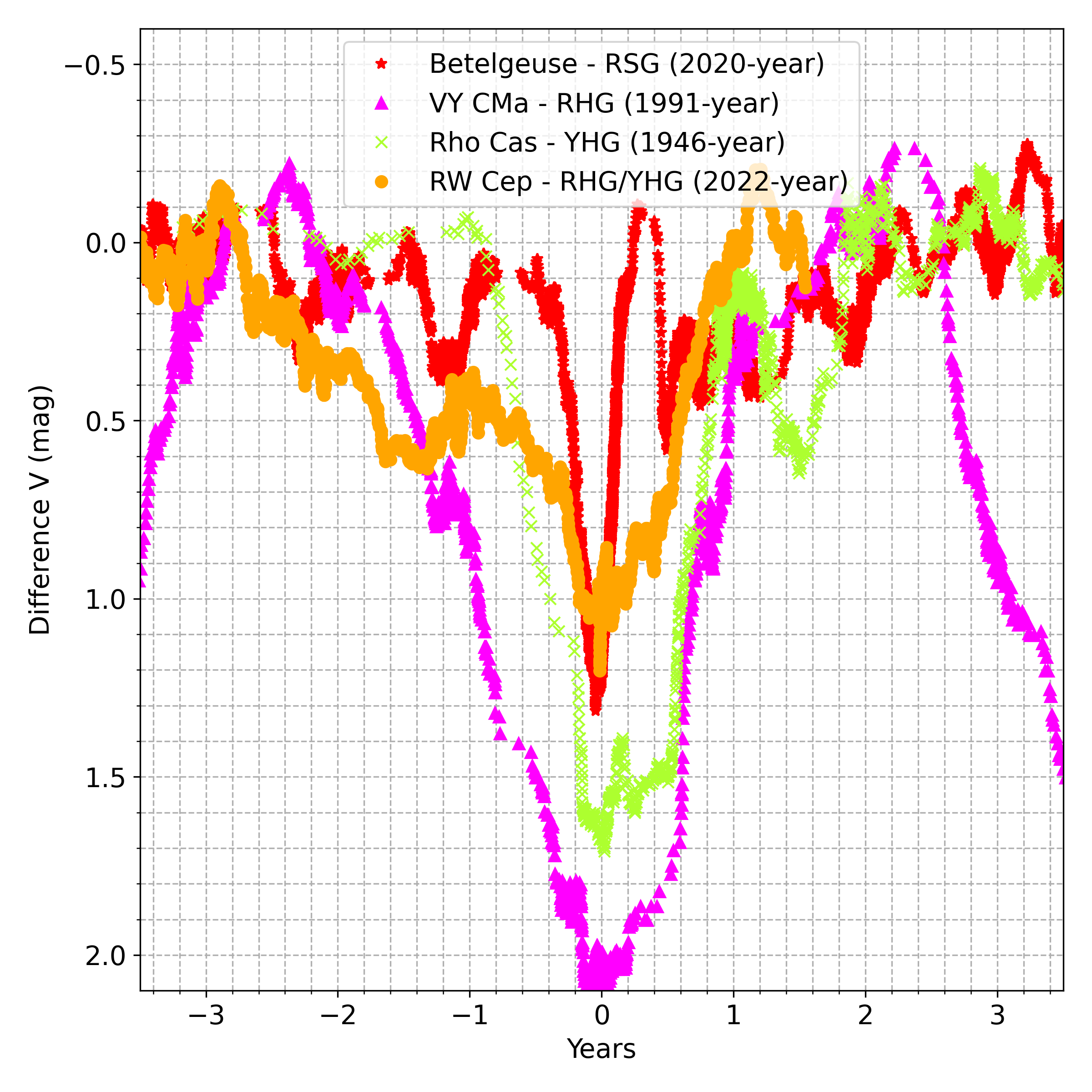}
\caption{
A comparison of the RW Cep 2020-2022 dimming event with the Great Dimming of Betelgeuse (recorded in 2019-2020), $\rho$~Cas (a major dimming event recorded in 1946-47), and VY CMa (a major dimming event recorded in 1988-1993). To compare the fall and rise times of the dimming events, the minimum point of each dimming event is set as the zero-point reference on the X-axis. We subtracted the mean value of each data set to obtain a relative magnitude along the Y-axis. This plot is made with AAVSO data.}
\label{Fig:compare_Betelgeuse_rhoCas}
\end{figure}

\section{Discussion}\label{Sec:Discussion}

\subsection{Revised physical sizes}

In Paper I, a distance range of 3.4 to 6.7 kpc was considered for RW Cep, with the possibility that it might be associated with the nearby Berkeley 94 star cluster, located approximately 6.5 arcminutes away. Recent studies by \citet{Delgado2013,Cantat-Gaudin2020,Almeida2023} estimated the age of the cluster to be between 13 and 32 Myr, approximately aligning with the evolutionary age of RW Cep. These studies also revised the distance to the cluster, providing estimates of $3900 \pm 110$pc \citep{Delgado2013}, 4097pc \citep{Cantat-Gaudin2020}, $3742\pm284$pc \citep{Almeida2023}, and 4000pc \citep{Hunt2024}. We adopted an average distance of $3935 \pm 152$~pc based on these estimates. However, we note that the Gaia EDR3 distance significantly diverges, with a value of $6.7^{+1.6}_{-1.0}$ kpc \citep{BailerJones2021}.

Taking into account the average Berkeley 94 star cluster distance and the post-dimming diameter of RW Cep measured at $2.60\pm0.01$ mas, we calculate the physical size of the stellar photosphere of RW~Cep to be $1100\pm44 R_\odot$. Furthermore, considering the angular diameter of the circumstellar envelope, measured at $4.3\pm0.2$ mas (see Section~\ref{Sec:pmoired_image} and Section~\ref{Fig:CO_lines_images}), we estimate its physical radius to be $1820\pm110 R_\odot$.


\subsection{A Dust Clump Hypothesis for the Dimming of RW~Cep}\label{Sec:SME}

A comparison of images of RW~Cep obtained during its 2022 December dimming phase (Paper I) and subsequent re-brightening (see Figure~\ref{Fig:compare_images_2022_2023}) reveals striking differences. We suggest that a dust clump obscured the west side of RW~Cep during this dimming event. Several lines of evidence support this hypothesis:

\begin{itemize}
    \item \textbf{Changes in stellar size:} We found that the limb-darkened photosphere diameter is $8\pm1$\% (Table~\ref{Table:diameter_fit}, Section~\ref{Sec:compare_images}) larger during the re-brightening phase than during the dimming phase. This substantial difference suggests that the stellar photosphere was partially blocked during the dimming phase, possibly by a dust clump, making the star appear smaller. The subsequent clearing of the dust allowed the star to return to its normal size.
    \item \textbf{Gas cloud outburst:} Concurrent H$\alpha$ observations in the visible wavelength spectrum by \citet{Kasikov2024} show a strong, blue-shifted emission during the dimming phase, suggesting a gas outburst directed towards us. This is consistent with the formation of a dust clump.
    \item \textbf{Dust evolution:} A visual inspection of our CHARA images taken in 2023 while re-brightening reveals small changes in images from one epoch to another during the star’s brightening, providing insights into the temporal evolution of the obstructing dust (see Figures~\ref{Fig:oitools_images}). 
    \item \textbf{Dust extinction:} The role of dust formation for the Great Dimming event is consistent with
changes recorded in the near-infrared spectrum.  We obtained near-IR spectra
of RW~Cep both during the the dimming and recovery stages with the
TripleSpec spectrograph on the 3.5 m telescope at Apache Point Observatory \citep{Wilson2004}.
The spectrum obtained during the re-brightening stage was made and reduced
in the same way as the first spectrum (as described in Paper I).  It is
difficult to obtain accurate absolute fluxes, given the extreme brightness
of the star and the finite slit size, so for the new observation we
renormalized the fluxes by comparing them to the H-band fluxes in a full-brightnss
spectrum of RW~Cep from the IRTF spectral library that was shown in Paper I.
This flux renormalization assumes that the star had returned to its full
brightness by the date of observation (2023 Sep 26; HJD 2460213.66). 
Inspection of Figure~\ref{Fig:dimming_light_curve} indicates that the re-brightening was not totally complete by this date, so we caution that the estimate of the flux in
the new spectrum may be slightly overestimated.  Figure~\ref{Fig:nir_spectrum} shows
a comparison of the spectrum during the Great Dimming (2023 January; black)
with that near full recovery (2023 September; gray).  We see that the
flux variations at longer wavelengths were minor, while large flux increases
occurred at shorter wavelengths during brightening.  Dust extinction is
much larger at shorter wavelengths, so these observations confirm the idea
that the Great Dimming event was caused by foreground dust formation that
blocked the flux from the star.
\end{itemize}

We propose a similar explanation offered for the Betelgeuse dimming. The explanation for the origin of the Betelgeuse Great Dimming involves the formation of a rare and hot convective bubble on the surface of star, accumulating material propelled by the creation of a large plasma outflow due to powerful deep successive shocks \citep{Kravchenko2021} beneath the stellar photosphere \citep{Dupree2020, Dupree2022}. This hot convection bubble transforms into a surface mass ejection driven by pulsation and magnetic fields \citep{Dupree2022, Humphreys2022}. The ejected gas cloud moves away from the star, cools down, and gradually forms dust in front of the star. The opaque dust obstructs the star in the southern part, causing the star to dim \citep{Montarges2021}. 
Due to the loss of material from the hot spot region, a cooler area emerges on the photosphere, appearing as a dark spot in the line of sight. As the dust next condenses into larger grains, the opacity drops, and the star returns to its normal brightness.

We suggest that similar processes might have been triggered in RW~Cep. Gas ejection began in 2020 in our direction, expanded outwards, and cooled to begin forming dust, blocking the western side of the star with a projected size of $0.45 R_*$.
This caused RW~Cep to appear darker during its dimming in  2022 December. As the dust moved further away from the star, it cleared from our line of sight, causing the star to reappear in its normal shape in the epochs of 2023. Additionally, the dust cloud is not entirely aligned in our direction, so as it starts to move off the face of the star, this cloud motion out of the line of sight to the star could happen rapidly. Thus, the return to regular brightness is more rapid than the fading. This asymmetry in the light curve (slow fade, rapid re-brightening) appears in other stars too (Figure~\ref{Fig:compare_Betelgeuse_rhoCas}).

While the dimming of RW~Cep shares some similarities with Betelgeuse, it is important to note that there are also key differences. Figure~\ref{Fig:compare_Betelgeuse_rhoCas} compares the RW~Cep dimming event (2020-2022) to other massive stars experiencing significant mass loss: (i) the red supergiant (RSG) Betelgeuse Great Dimming (2019-2020), (ii) the red hypergiant (RHG) VY Canis Majoris major dimming event  \citep[1988-1993, ][]{Humphreys2021AJ....161...98H}, and (iii) the yellow hypergiant (YHG) $\rho$~Cas major dimming event \citep[1946-47, ][]{Beardsley1961ApJS....5..381B, Maravelias2022}.  To ensure consistency, the minimum of each event was utilized as the reference point for both relative magnitude and date. Notably, of the dimming event,  i.e., the fall and rise time of  RW Cep, lasted around 4 years, significantly longer than Betelgeuse ($<1$~y) and $\rho$~Cas ($\sim2$~y) but shorter than VY~CMa (over 5 years). The dimming depth of RW~Cep ($\sim1.2$ mag) resembled the Great Dimming of Betelgeuse, while $\rho$~Cas and VY~CMa dimmed further (more than 1.7 mag). These differences could potentially be attributed to the size of each star, as well as the size, opacity, and geometry of the dust clump blocking the star from our line of sight. If the dust clump takes longer to lose its opacity or to disperse by moving away from the star, the dimming dip duration will be prolonged. The depth of these dimming events is related to the size, density, and geometry of the dust clumps.

Table~\ref{Table:YHG_RWCep_compare} presents a comparison of properties of these stars. RW~Cep has an estimated radius $1100\pm44~R_\odot$, making it approximately twice the size of $\rho$~Cas, which ranges from $500-760~R_\odot$ \citep{Anugu2024}, and 1.25 times larger than Betelgeuse, with a radius ranging from $750-1000~R_\odot$ \citep{Joyce2020, Kravchenko2021}. This prolonged dimming duration for the larger stars, VY~CMa \citep[$1420\pm120~R_\odot$, ][]{Wittkowski2012} and RW~Cep might imply that these stars have experienced greater mass loss and have a larger amount of dust surrounding them, potentially leading to an extended period of obscuration. Furthermore, if the outflow velocities are comparable, it takes longer for the material to travel due to the larger dimensions of these stars.
One exception is $\rho$~Cas, which is the smallest star in our sample but experienced a longer dimming duration ($\sim2~y$) than Betelgeuse ($<1~y$). This suggests that the ejected material from $\rho$~Cas may have different dust formation properties, such as grain size and chemistry, which obscure the star for a longer period.

Since RW~Cep (YHG) and Betelgeuse (RSG) belong to different object classifications, potential differences in dust chemistry and grain size might complicate our understanding of the RW~Cep dimming phenomenon.  A detailed analysis of dust properties of RW~Cep is beyond the scope of this current work.

Another striking difference between Betelgeuse and RW~Cep is the temporary change in pulsation period in post-dimming. After the Great Dimming of Betelgeuse, the pulsation period was observed to have halved from its 420-day primary period \citep[e.g.,][]{MacLeod2023ApJ...956...27M}.  Hydrodynamic simulations by \cite{MacLeod2023ApJ...956...27M} suggest that a significant upwelling of a hot convection bubble can disrupt the phase coherence of the star's pulsation, leading to an expansion of the surface while deeper layers contract. Consequently, this led to a change in the pulsations from the 420-day pulsation period to the 200-day. They also suggest that Betelgeuse will likely return to its previous 420-day fundamental mode pulsation within the next 5-10 years. As of the time of writing, RW~Cep has fully recovered to its typical brightness. 
However, since March 2024, it started dimming again (see Figure~\ref{Fig:dimming_light_curve}). Future observations may or may not reveal changes in the  pulsation period of RW~Cep (refer to Section~\ref{Sec:aavso_light_curve}). 

Similar to the 2022 dimming episode, the other major dimming events of RW~Cep recorded in 1908.5, 1936, 1950.5, 1962, and 1988 —characterized by a more significant drop in visual magnitude and a longer duration ($>3$~y) — might have been triggered by a hot convection cell-driven mass loss mechanism (see Figures~\ref{Fig:dimming_light_curve_total} and \ref{Fig:dimming_light_curve}). Apart from these major dimming events, the more frequent dimmings, which exhibit a lower drop in visual magnitude and shorter duration ($1-1.5$~y), are likely driven by pulsation-enhanced dust formation above the photosphere.

\begin{deluxetable*}{c c c c c c c c c c c}
\tablecaption{Comparing RW~Cep with Red and Yellow hypergiants}
\label{Table:YHG_RWCep_compare}
\tablewidth{0pt}
\tablehead{
\colhead{Object} & Class & \colhead{Init Mass} & \colhead{$T_{\rm eff}$} &  \colhead{$\log(L/L_\odot)$} & \colhead{$R/R_\odot$} & \colhead{Spectral} & \colhead{Mass Loss Rate} \\
\colhead{} & & \colhead{$M_i/M_\odot$} & \colhead{K} &  \colhead{} & \colhead{} & \colhead{Classification} & \colhead{$M_\odot {\rm yr}^{-1}$ }
}
\startdata
VY CMa & Red & $30$-$40$ [1]  &  $3490\pm90$ [2] & $5.4\pm0.1$ [2] & $1420\pm120$ [2] & M5e Ia [2] & $10^{-3}$-$10^{-4}$ [4] \\
RW~Cep    & Red/Yellow &25-40   & $4200–4400$ [3,5]   & 5.47-5.73 [3,4] &  $1100\pm44$ [6] & K2Ia-0 - M2Ia-0 [7,8] & $\sim7\times10^{-6}$ [3]\\
$\rho$~Cas   & Yellow & 40 [9] & $4500$-$8000$ [9,10]  & $5.48\pm0.32$ [10] & 564-700 [17] & F8 - G2 Ia0 [9] &  $10^{-3}$-$10^{-6}$ [11]  \\
IRC+10420 & Yellow & $20$-$40$ [13] & $7930\pm140$ [14] & 5.42-5.60 [14,15] & $442 \pm 84$ [18] & F8Ia+ - A2I [15,16] & $2\times10^{-3}$  [4]\\
HR~8752   & Yellow & 20-30 [14] & $4000$-$8000$ [14,21]  & 5.43 [10] & $511 \pm 112$ [19]  &  K5Ia0 - A6Ia+ [14] & $10^{-6}$ [14]\\
HR~5171~A & Yellow & 25 [12] & $4287\pm760$ [12] & 5.79 [12] & $1575 \pm 400$ [12]  &  G0/2-K1/3Ia+ [12] & $7\times10^{-6}$ [20]\\
\enddata
\tablecomments{1. \citet{Humphreys2007}, 2. \citet{Wittkowski2012}, 3. \citet{Jones2023}, 4. \citet{Shenoy2016},  5. \citet{Anugu2023}, 6. This work, 7. \citet{Keenan1989}, 8. \citet{Watson2006},  9. \citet{Kraus2019},  10. \citet{vanGenderen2019}, 11. \citet{Lobel2003}, 12. \citet{Wittkowski2017}, 13. \citet{Clark2014}, 14. \citet{Nieuwenhuijzen2012}, 15. \citet{Klochkova2019},  16. \citet{Klochkova2002}, 17. \citet{Anugu2024}, 18. \citet{Oudmaijer2013},  19. \citet{vanBelle2009},    20. \citet{Chesneau2014}, 21. \citet{Kasikov2024b}. We used the Gaia DR3 distances \citep{BailerJones2021} in converting the angular diameters into physical radii. } 
\end{deluxetable*}

\subsection{Evolutionary State of RW~Cep: Yellow hypergiant?}

A crucial first step in comparing the RW Cep dimming  event to similar stellar phenomena is precisely classifying the star.  The exact classification of RW Cep as a red or yellow hypergiant remains unclear \citep{Castro-Carrizo2007A&A...465..457C, Jones2023}. The red or yellow hypergiants are a small class of evolved massive stars that represent a short-lived evolutionary phase near the top of the Hertzsprung-Russell diagram. They are extremely rare, highly unstable, and experience severe mass loss. Red hypergiants are the most luminous cool hypergiants, while YHGs are intermediate temperature or warm hypergiants. An example of a red hypergiant (RHG) star is VY~CMa \citep[][references within]{Humphreys2021AJ....161...98H}.

Yellow hypergiants (YHG) may represent the transition from red supergiant objects that evolve bluewards in the Hertzsprung-Russell (H-R) diagram. However, the details of such evolution are still poorly understood \citep{deJager1998, Nieuwenhuijzen2000}.  They exhibit strong spectral variability and recurrent eruption events, with large mass loss rates of up to $\sim5\times10^{-2} M_\odot{\rm yr^{-1}}$ when they experience an outburst \citep{Lobel2003}.  Most YHGs have a temperature near the cool border of the yellow void, an unstable region with effective temperatures between $\sim7000-12000$ K. Four well-studied YHGs are $\rho$~Cas \citep{Lobel2003, Kraus2019, Anugu2024}, IRC+10420 \citep{Humphreys2002AJ....124.1026H, Tiffany2010AJ....140..339T}, HR 8752 \citep{Nieuwenhuijzen2000, Nieuwenhuijzen2012, Kasikov2024b}, and HR 5171 A \citep{Chesneau2014, Wittkowski2017}.

Table~\ref{Table:YHG_RWCep_compare} presents a comparison of the properties of RW~Cep with those  of established RHG and YHGs. This comparison aims to determine if RW~Cep can be classified as a YHG or RHG.

\begin{itemize}
\item Luminosity: Estimated luminosity of RW~Cep $\log(L/L\odot)$ = 5.47 - 5.73 falls within the range established for YHGs $\log(L/L\odot)=5.4-5.8$ \citep{Humphreys1978ApJS...38..309H, deJager1998, Jones2023}. This is significantly lower than the typical luminosity of red hypergiants.
  \item Temperature: With an effective temperature ranging from 4,400 K to 5,018 K \citep{Jones2023, Meneses-Goytia2015A&A...582A..96M}, RW~Cep falls on the lower end of the YHG range $T_{\rm eff} = 4000-7500$ \citep{Nieuwenhuijzen2000,Nieuwenhuijzen2012}. 
  \item Episodic Outbursts and Circumstellar Envelope Formation: A defining feature of YHGs is episodic outbursts that lead to mass loss and the formation of a circumstellar envelope and shells. Similar to the well-known YHG $\rho$~Cas, RW~Cep exhibits outbursts visible as sudden dips in brightness followed by rebrightening, as evidenced by its light curve data since 1900 (see Figure~\ref{Sec:aavso_light_curve}). Spectral studies also reveal the presence of a moving gas shell around RW~Cep, similar to $\rho$ Cas \citep{Merrill1956}.
  \item Spectral Variation: Another characteristic of YHGs is a change in spectral type over time \citep{Nieuwenhuijzen2012}. The spectral classification of  RW~Cep is poorly studied over the years. However, it seems to have shifted from G8-Ia \citep{Morgan1950ApJ...112..362M} to K2-Ia  \citep{Rayner2009ApJS..185..289R,Messineo2021AJ....162..187M} and finally to M2-Ia \citep{Humphreys1978ApJS...38..309H, Watson2006, Skiff2010yCat....102023S}.
\end{itemize}

The combination of its luminosity, temperature variations, episodic outbursts, mass loss, presence of a moving gas shell, spectral variations, and the presence of circumstellar envelope and stellar surface bright and dark features bear many similarities to the well-studied YHG $\rho$~Cas \citep[]{Lobel2003,Kraus2019,Anugu2024}. These factors  suggest that RW~Cep is a YHG candidate and can be considered an analogue to $\rho$~Cas.

\paragraph{Similarities and Differences: RW~Cep vs. yellow hypergiant $\rho$~Cas } 
To gain a deeper understanding of RW~Cep, we present a comparison with $\rho$~Cas.  Both stars are surrounded by shells of gas and dust, as evidenced by CO gas emission \citep[]{Lobel2003,Anugu2024}. A striking similarity is that the both stars experience episodic mass loss, but  RW~Cep, appears to be more active compared to $\rho$~Cas. It experiences outbursts roughly every 15 to 35 years, compared to a  frequency of 20 to 50 years $\rho$~Cas. Since 1900, RW~Cep has undergone six estimated outbursts (see Figure~\ref{Fig:dimming_light_curve_total}), while $\rho$~Cas has only had four since 1940 \citep{Maravelias2022}. Despite this difference in outburst frequency, both stars share a similar pulsating nature, exhibiting semi-regular variations in brightness. RW~Cep pulsates every 350 to 800 days (see Figure~\ref{Fig:dimming_light_curve_period}), while $\rho$~Cas pulsates every 320 to 500 days \citep{vanGenderen2019}. The amplitude of these brightness changes due to pulsation is also comparable, with RW~Cep varying by about $\pm0.3$ magnitudes and $\rho$~Cas by $\pm0.2$ magnitudes (in the V-band). Interestingly, both stars exhibit a similar light curve pattern, characterized by an increase in V-band brightness before an outburst. \citet{Lobel2003} documented this trend in $\rho$~Cas prior to the Millennium outburst that occurred in 2000-2001, and a similar trend has been observed in RW~Cep (see Figure~\ref{Fig:dimming_light_curve_total}). Finally, both these objects show blue-shifted strong H$\alpha$ emission during the dimming compared to before and after the dimming \citep{Lobel2003,Kasikov2024}.

There are a few differences. Based on their effective temperatures, $\rho$~Cas, with $T_{\rm eff}=4500-8000$K \citep{Kraus2019,vanGenderen2019}, appears to be moving towards the blueward loop closer to the low-temperature border of the yellow void in the HR diagram \citep{deJager1998}. In contrast, RW\,Cep, with  $T_{\rm eff}=4200-4400$~K \citep[Paper I,][]{Jones2023}, lies at the lower end of the yellow hypergiant temperature range, which spans from 4000-7000~K. The effective temperature changes during dimming cycles differ significantly. RW~Cep cooled by a smaller amount (300~K) during its most recent outburst \citep[2022-2023, ][]{Anugu2023} compared to $\rho$~Cas much larger drop of 3000~K during its 2000-01 outburst \citep{Lobel2003}. The spectral variations in $\rho$~Cas are clearly evident, changing from F8 to G2 Ia0. Similarly RW~Cep also exhibits spectral variations.

\section{Summary and Conclusions}\label{Sec:Summary}

This study investigates the 2022 dimming and subsequent brightening of the massive star RW~Cep. We conducted observations of RW~Cep using the world's largest interferometer at H and K-band wavelengths. Images of the star were reconstructed from the interferometric data using a model fitting and three different image reconstruction methods to check the robustness of the images. The resulting images are mostly in agreement, confirming the validity of the images.

During the dimming phase in  2022 December, the CHARA images recorded an asymmetric shape, with a darker western side. As the star entered the re-brightening phase in 2023, the western side reappeared brighter again in the CHARA images. Furthermore, observations from 2023 July, August, September, and October reveal changes from epoch to epoch. These changes could be attributed to the time-varying dispersal of dust away from the star. Finally, we also found that the star's photosphere diameter increased by $8\pm1$\% during the re-brightening phase compared to the dimming phase. This substantial increase suggests that the stellar surface was partially blocked during the dimming phase, possibly by a dust clump. The star appeared smaller and asymmetrical during the Great Dimming compared to the 2023 re-brightening images, indicating that some clearing out has occurred.

We propose that the dimming in RW~Cep is due to a recent surface mass ejection event, similar to processes observed in Betelgeuse. This ejected material likely cooled and formed a dust cloud that partially blocked the starlight in our line of sight, causing the observed dimming. As the dusty ejecta expand and clear, the star re-brightened, revealing the previously obscured western region. While the dust formation scenario offers a compelling explanation for the Great Dimming of RW~Cep, several questions remain unanswered. A more comprehensive understanding requires a detailed analysis of the dust properties around the star, including composition, grain size distribution, and total mass. Alternative explanations, such as the formation of giant cool spots on the stellar surface causing the observed changes, are potentially ruled out. The strong evidence from the H$\alpha$ emission and blueward shifts of the gas cloud during the Great Dimming \citep{Kasikov2024} supports the conclusion that a mass ejection occurred in our direction and that dust formation in the gas clump caused the fading. 

Overall, this study suggests that RW~Cep is a yellow hypergiant candidate star experiencing episodic dimming events caused by potential dust formation around the star. 

Since March 2024, the star has started dimming again. A follow-up program involving spectroscopic, photometric, and imaging observations of RW~Cep could reveal outburst events, significant variability in effective temperature, large changes in mass loss rates, and provide essential information about the fundamental properties of atmospheric dynamics and wind physics. 

We interpret the bright spots on the stellar surface with contributing fluxes of 4\%, 8\% and 9\% (Sections~\ref{Sec:pmoired_image} and \ref{Sec:image_recon_conti}) observed during the RW~Cep re-brightening as giant convection cells on the stellar surface. We plan to use the CHARA Array to study the dynamics of these convection cells over time, which could reveal a link between the convection cells and long secondary period variability. The long secondary periods maybe linked to the turnover duration time of large convection cells on the stellar surface, which are recognized for their substantial depth and vigorous convection \citep{Stothers1971A&A....10..290S,Stothers2010ApJ...725.1170S,Stothers2012}. Long-term CHARA monitoring, coupled with AAVSO light curve data, may help establish connections between convection cell activity, pulsations, long secondary periods, and mass loss events.

\vspace{0.5 cm}
{
We would like to express our sincere gratitude to the anonymous reviewer for their insightful comments and constructive suggestions, which have significantly improved the quality of our manuscript.
We thank Jean-Baptiste Le Bouquin for his significant contributions to the development of the MIRC-X and MYSTIC instruments.
We are grateful to Naomi Elmberg for assistance in obtaining the DASCH photometry data for RW Cep. 
We thank Stephen Ridgway for his comments, which enabled improvement of the paper. 
This work is based upon observations obtained with the Georgia State University 
Center for High Angular Resolution Astronomy Array at Mount Wilson Observatory.  
The CHARA Array is supported by the National Science Foundation under Grant No.\
AST-1636624, AST-1908026, and AST-2034336.  Institutional support has been provided 
from the GSU College of Arts and Sciences and the GSU Office of the Vice President 
for Research and Economic Development.
We acknowledge with thanks the variable star observations from the AAVSO International Database contributed by observers worldwide and used in this research.
The work is also based on observations obtained with the Apache Point Observatory 3.5-meter telescope, which is owned and operated by the Astrophysical Research Consortium.
R.M.R. acknowledges support from the Heising-Simons Foundation 51 Pegasi b Fellowship.}
M.M. acknowledges funding from the Programme Paris Region fellowship supported by the R\'egion Ile-de-France. This project has received funding from the European Union’s Horizon 2020 research and innovation program under the Marie Sk\l{}odowska-Curie Grant agreement No. 945298.
J.D.M.\ acknowledges funding for the development of MIRC-X (NASA-XRP NNX16AD43G, 
NSF AST-1909165) and MYSTIC (NSF ATI-1506540, NSF AST-1909165).
F.B.\ acknowledges funding from the National Science Foundation under Grant No.\ AST-1814777.
S.K. acknowledges support from the European Research Council through a Starting Grant (Grant Agreement No. 639889) and Consolidator Grant (Grant Agreement No. 101003096), and STFC Consolidated Grant (ST/V000721/1).

\facility{CHARA, AAVSO}
\software{Astropy \citep{2022ApJ...935..167A},  PMOIRED \citep{antoine_merand_2024_10889235},  ROTIR \citep{Martinez2021}, OITOOLS \citep{Norris2021}, SURFING \citep{Roettenbacher2016}, mircx\_pipeline \citep{le_bouquin_2024_12735292}}

\appendix

\section{The CHARA Observation Log and Fitting Residuals}\label{Sec:Appendix_log_fit}

This section presents the CHARA observing log, the (u,v)-coverage of our observations, and the fitting residuals. The detailed observing log for the MIRC-X and MYSTIC observations is provided in Section~\ref{Table:RW_Cep_observations}. 

Figure~\ref{Fig:uv_coverage} presents the $(u,v)$-coverage of the 2023 observations. These $(u,v)$ coordinates specify which components of the Fourier transform of the sky brightness are represented in the visibilities and closure phases. The effective resolution of the interferometric reconstructed images is determined by the maximum extent of the $(u,v)$-coverage. Gaps in $(u,v)$-coverage can introduce artifacts in synthesized images. Our observations present good $(u,v)$-coverage, providing an opportunity to reconstruct reliable images. We combined the nights observed in 2023 July (nights 21, 22, and 23) into a single dataset  for a better $(u,v)$-coverage.  Similarly, we grouped the observations from 2023 September (nights 01 and 17) together, assuming that the stellar surface did not change on such short time sales. 

Figure~\ref{Fig:v2_t3} shows the visibilities $V^2$ and closure phases ($T3PHI$) of RW~Cep  in the K band. The CO bandhead  lines (2.3 $\mu$m) show lower $V^2$ compared to the continuum, indicating the presence of a circumstellar envelope. We also conducted medium and high spectral resolution observations ($\mathcal{R}=981$ and $\mathcal{R}=1724$) with the MYSTIC instrument. These observations aimed at characterizing the circumstellar envelope in the CO bands, and the results will be presented in a separate paper. 

Figure~\ref{Fig:pmoired_mystic_residuals}  presents the K-band ($2.0-2.39~\mu$m) fitting residuals from the PMOIRED model fitting for the epoch of 2023 July 21. The fitting residuals for the OITOOLS image reconstruction of the continuum are shown in  Figure~\ref{Fig:oitools_fit_mystic}. The residuals for other ROTIR and SURFING reconstructions are qualitatively similar.

\begin{deluxetable*}{l l l l l l}
\tablecaption{\label{Table:RW_Cep_observations}RW\,Cep observations from the CHARA Array instruments MIRC-X (H-band) and MYSTIC (K-band). The instrument configuration with spectral dispersion ($\mathcal{R}$) and calibrators are included. The calibrator star diameters adapted from the JMMC catalog~\citep{Bourges2017}. UT indicates the universal time at the beginning of the data recording. }
\tablewidth{0pt}
\tablehead{
 \colhead{DATE } & Instrument & Spec. Res. Power & \colhead{Calibrators} & \colhead{Diameter}  \\
 \colhead{(UT)}  &            & ($\mathcal{R}=\lambda / \triangle \lambda$) & \colhead{} & \colhead{(mas)} 
}
\startdata
2022-12-23$^{a}$ & MIRC-X  &50 &  HD 224870   & $0.653 \pm 0.053$  \\
                  & MYSTIC &100 &  HD 5916   & $0.527 \pm 0.037$\\
2023-07-21$^{b}$ & MIRC-X  &50 &  HD 224870   & $0.653 \pm 0.053$  \\
            & MYSTIC   &278 & HD 5916   & $0.527 \pm 0.037$ \\
2023-07-22$^{b}$ & MIRC-X  &50 & HD 216206   & $0.885 \pm 0.074$ \\
            & MYSTIC   &1724 & HD 219080   & $0.692\pm0.067$  \\
2023-07-23$^{b}$ & MIRC-X  &50  & HD 216206   & $0.885 \pm 0.074$  \\
            &  MYSTIC   &981 & HD 217694   & $0.897\pm0.058$\\
2023-08-04 & MIRC-X  &190  & HD 219080   & $0.692\pm0.067$ \\
            & MYSTIC   &278 & HD 224870   & $0.653 \pm 0.053$ \\
            & & & HD 16780   & $0.940  \pm  0.090$  \\
2023-09-01$^{c}$ & MIRC-X  &50  & HD 218187   & $0.622 \pm 0.015$  \\
            & MYSTIC   &50 & HD 26553   & $0.500\pm 0.045$ \\
& & & HD 211822   & $0.563\pm 0.013$\\
& & & HD 211982    & $0.569\pm0.014$ \\
2023-09-17$^{c}$ & MIRC-X  & 190 & HD 219080 & $0.692\pm0.067$\\
                 & MYSTIC  & 278 & HD 211982 & $0.569 \pm 0.014$  \\
& & & HD 211822  & $0.563 \pm 0.013$ \\
2023-10-25 & MIRC-X  &190 & HD 219080   & $0.692\pm0.067$ \\
            & MYSTIC   &278 & HD 211982      & $0.569 \pm     0.014$  \\
& & & HD 6282   & $0.554 \pm        0.012$
\\\enddata
\tablecomments{Observing date with flag $a$ is  already published in Paper I. Observing dates with flag $b$ are combined together in a single group to achieve better $(u, v)$-coverage. Same for $c$ flagged nights.} 
\end{deluxetable*}

\begin{figure*}[h!]
\centering
\includegraphics[width=\textwidth]{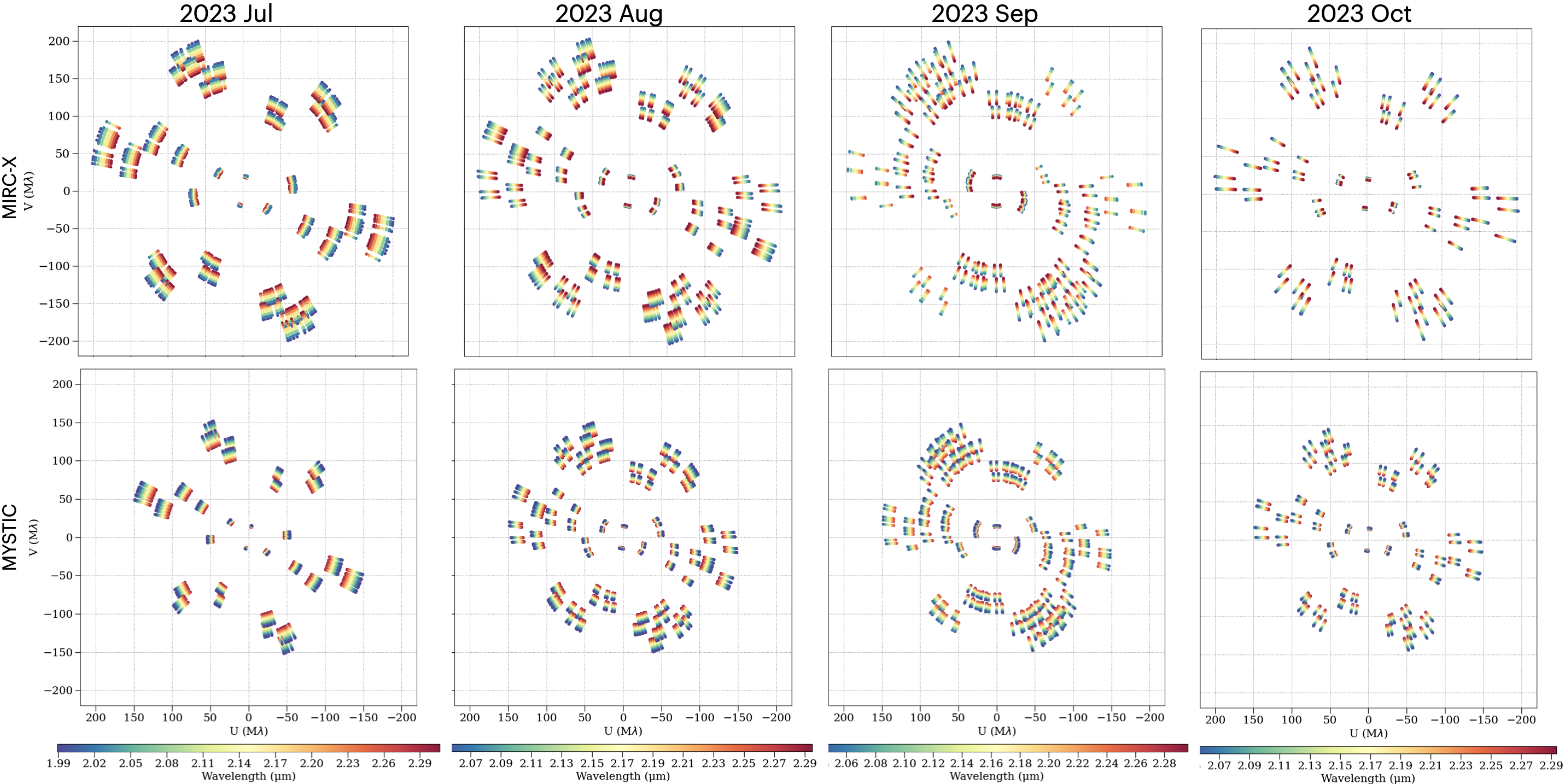}
\caption{
The $(u,v)$-coverage of RW~Cep observations for different epochs. The top panel is for MIRC-X and lower panel is for MYSTIC. The $(u,v)$-coordinates represent the projections of the baseline vectors on the plane of the sky of the target.}
\label{Fig:uv_coverage}
\end{figure*}

\begin{figure*}[h!]
\centering
\includegraphics[width=0.8\textwidth]{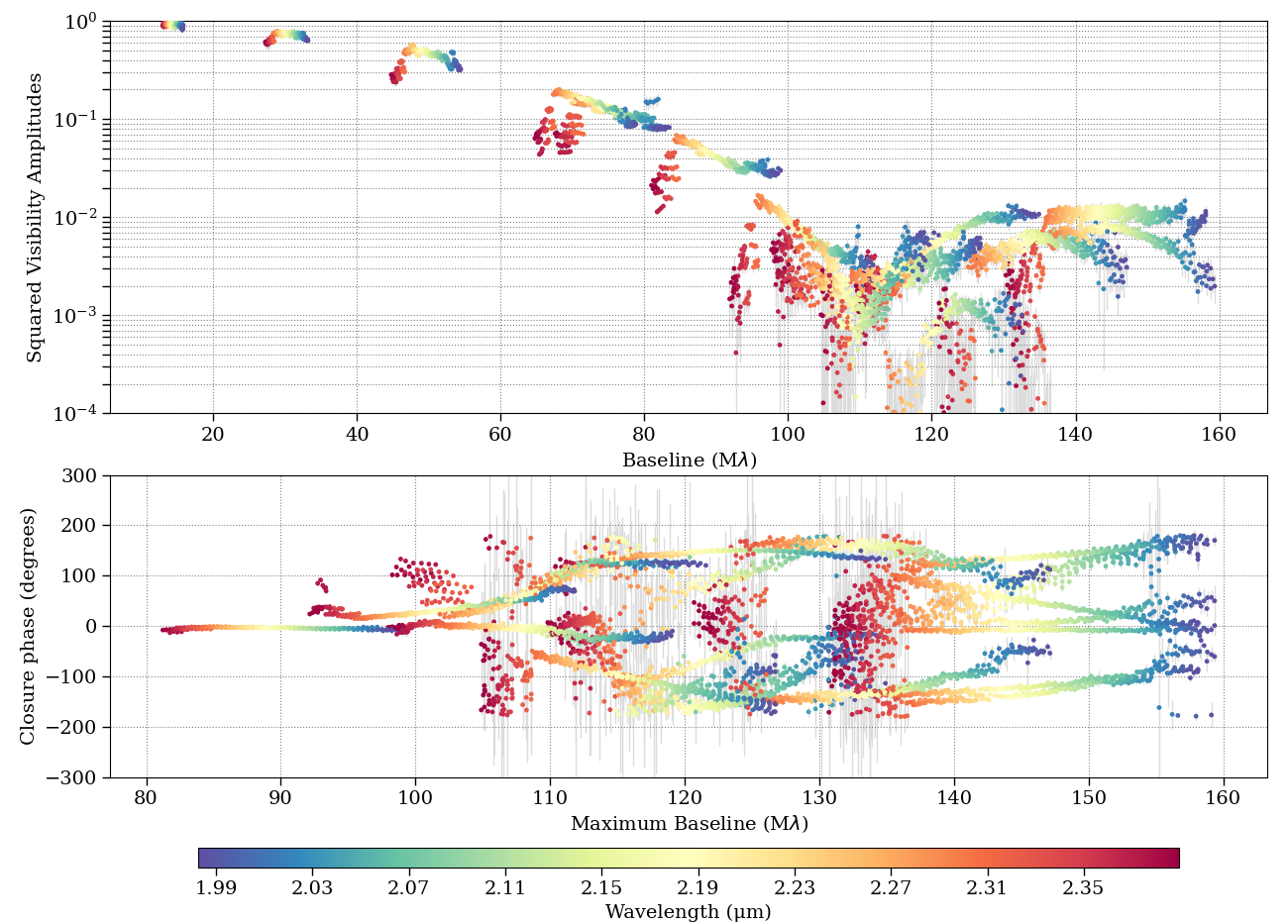}
\caption{The K-band observations of RW~Cep from 2023 July 21. The squared visibility and closure phases are presented in the top and bottom panels, respectively. The color bar represents the wavelength. The squared visibility in the CO bandhead  lines ($2.3~\mu$m) is smaller, indicating a larger angular size  compared to that for the continuum.
}
\label{Fig:v2_t3}
\end{figure*}

\begin{figure*}[h!]
\centering
\includegraphics[width=0.9\textwidth]{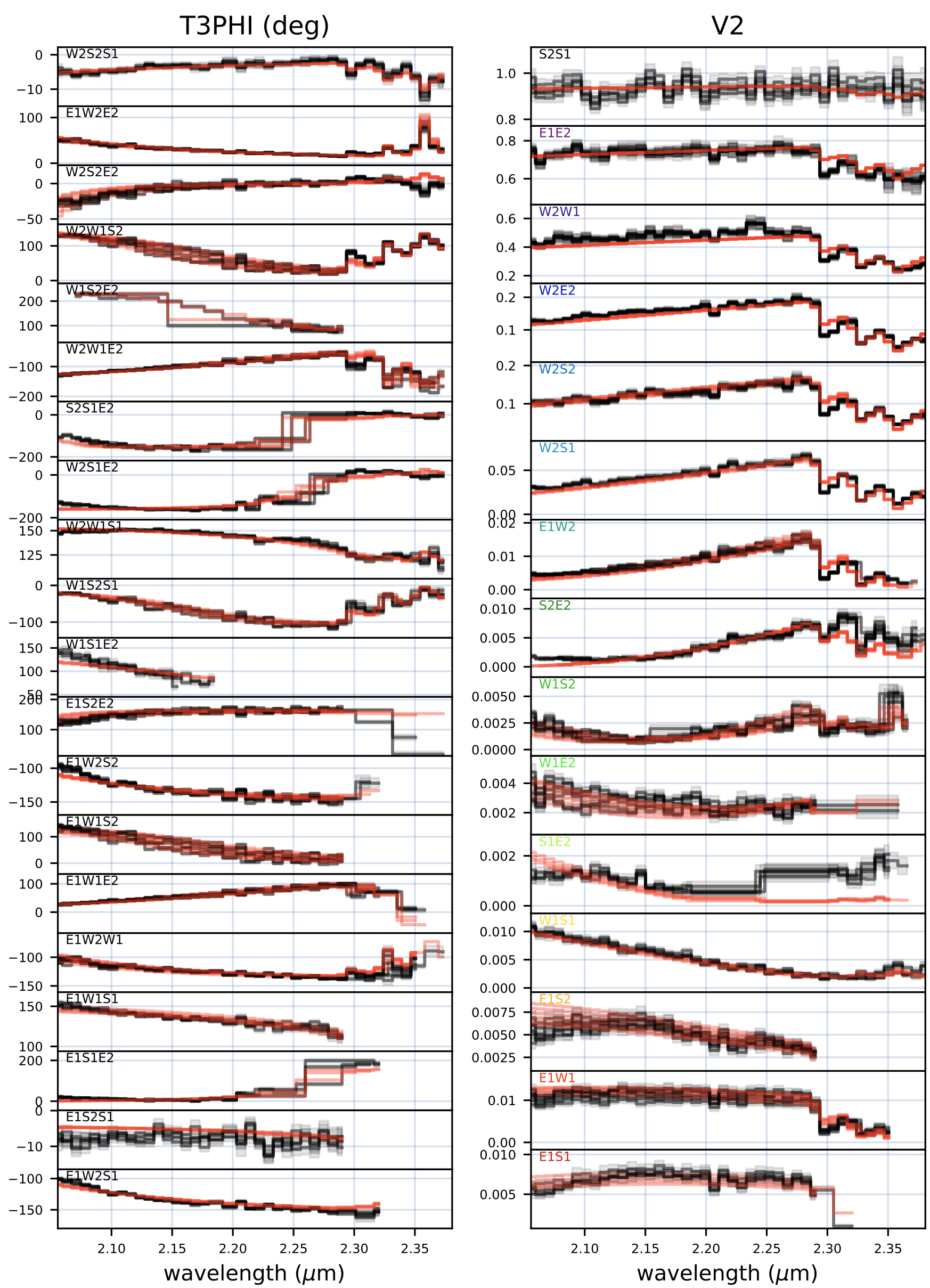}
\caption{The PMOIRED fit residuals ($V^2$ and $T3PHI$) of the MYSTIC K-band data ($\lambda = 2.0-2.39~\mu$m) for the epoch 2023 Jul 21 are shown. The black lines represent the interferometric data, while the red lines depict the analytical model fits.} The CHARA baselines are indicated in each baseline box. For details on the PMOIRED model fitting procedures, refer to the Section~\ref{Sec:pmoired_image}.
\label{Fig:pmoired_mystic_residuals}
\end{figure*}

\begin{figure*}[h!]
\centering
\includegraphics[width=\textwidth]{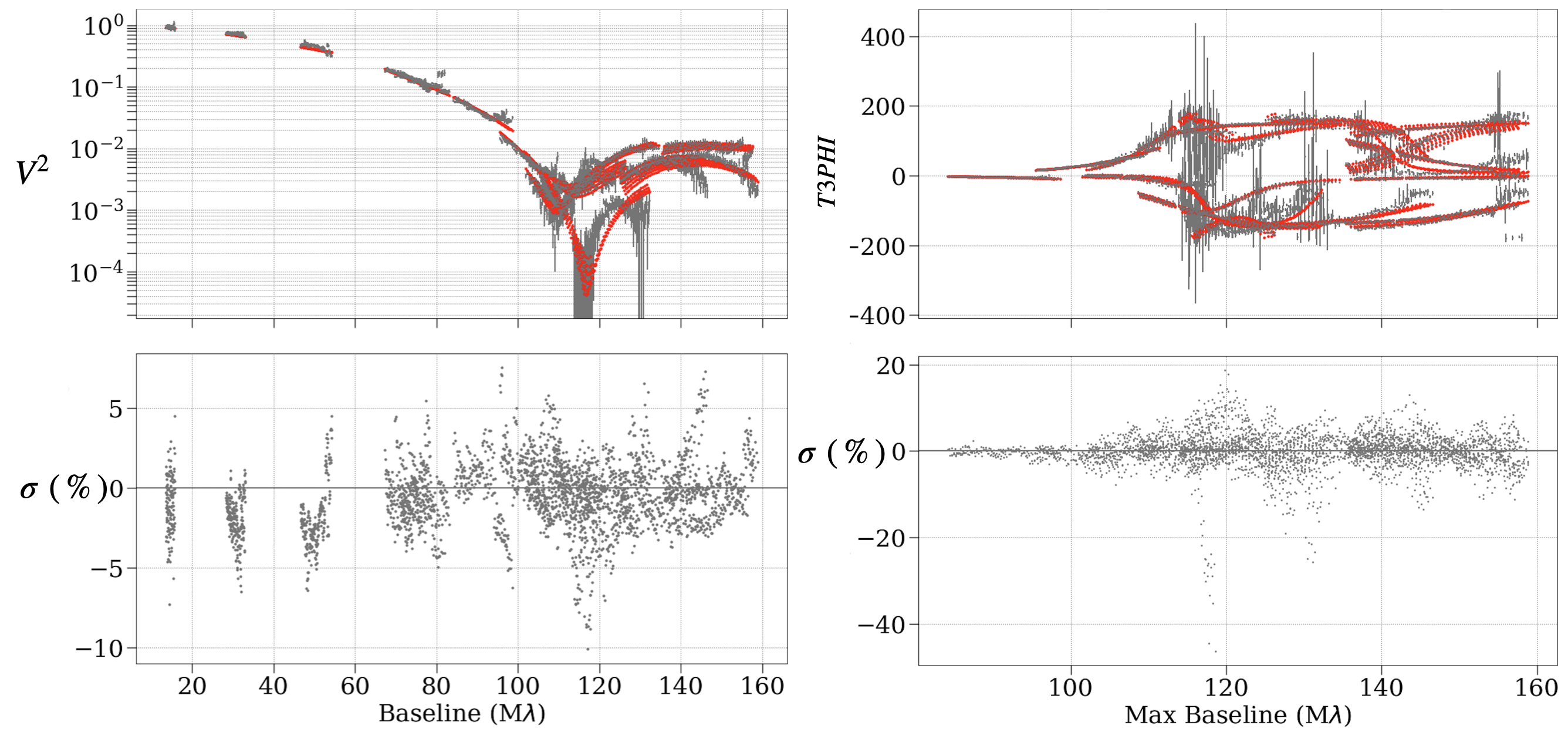}
\caption{The OITOOLS fitting residuals for the $K$-band ($2.0-2.31~\mu$m) visibilities (left) and closure phases (right). These are for the epoch 2023 Jun 21. 
The grey symbols with vertical error bars are the measurements
and the red dots represent the corresponding fits from the OITOOLS reconstructed images. 
The lower panels show the fitting residuals in standard deviation ($\sigma$~\%). See Section~\ref{Sec:image_recon_conti} for the image reconstruction procedures used.   
}
\label{Fig:oitools_fit_mystic}
\end{figure*}

\section{Additional Image Reconstructions with ROTIR and SURFING}\label{Sec:ROTIR_SURFING}

We used two additional image reconstruction software packages to reconstruct images from the interferometric datasets: (i) ROTational Image Reconstruction (ROTIR) \citep{Martinez2021}\footnote{\href{https://github.com/fabienbaron/ROTIR.jl.git}{https://github.com/fabienbaron/ROTIR.jl.git}}, and
(ii) SURFace imagING \citep[SURFING, ][]{Roettenbacher2016}. ROTIR and SURFING make use of HEALPix tessellation \citep{Gorski2005ApJ...622..759G} to generate three-dimensional surfaces of stars. They assign a specific intensity to each element on the three-dimensional surface of the star. ROTIR is coded in the Julia programming language and uses OITOOLS libraries, whereas SURFING is developed using IDL. 

The ROTIR software allocates flux to surface patches on a rotating star, but in this case only the visible hemisphere is considered because the rotational period is much longer than the span of the observations. We used total variation regularization with hyperparameter weight 0.05. The SURFING algorithm iteratively solves the brightness distribution on the pixels of the sphere. Our ROTIR and SURFING images consist of approximately 22 pixels across the stellar diameter.  Figure~\ref{Fig:ROTIR_images} and ~\ref{Fig:SURFING_images} show the ROTIR and SURFING reconstructed images. These image reconstructions agree well with the Figures~\ref{Fig:pmoired_images_all_2023_epochs} and \ref{Fig:pmoired_images_compare_2022_2023} obtained from the model fitting (see Section~\ref{Sec:pmoired_image}). The fitting residuals for the OITOOLS image reconstruction of the continuum are shown in Appendix~\ref{Sec:Appendix_log_fit}. The residuals for other ROTIR and SURFING reconstructions are qualitatively similar.

Upon visually examining the continuum wavelength images produced by the OITOOLS, ROTIR and SURFING software, we observe a remarkable agreement among the images, with some noteworthy differences. A significant distinction between the OITOOLS and ROTIR and SURFING images are the assumed shape of the star. Especially in the SURFING images, where the bright and dark spots appear to be larger, although the pixel scales are similar in scale to ROTIR. A known issue with the current version of the ROTIR software is its incorrect handling of limb-darkening effects. The OITOOLS images, made with fewer assumptions, present a star with an asymmetric shape, while ROTIR and SURFING assume a spheroid-shaped star. However, considering the dark spots on ROTIR and SURFING images in the South-East and North-East, the overall shape of the star appears similar to the images generated by the OITOOLS software. Given the asymmetric shape of the star in the presence of the circumstellar envelope (see Figure~\ref{Fig:CO_lines_images}), there is greater confidence in the reliability of the images generated by the OITOOLS software compared to those made with ROTIR and SURFING.

\begin{figure*}[h!]
\centering
\includegraphics[width=\textwidth]{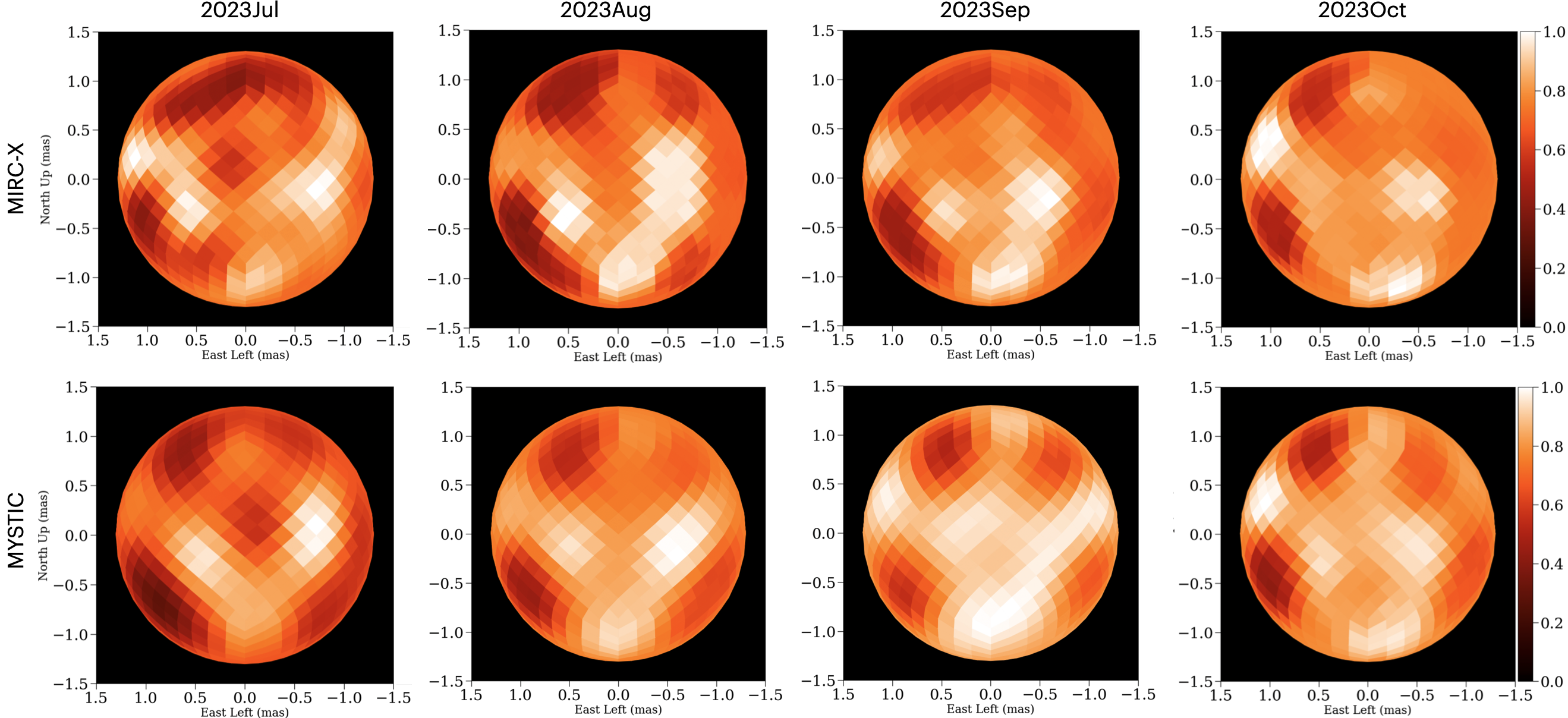}
\caption{The ROTIR reconstructed images of RW~Cep for epochs 2023 July, August, September and October. The top panel corresponds to the H-band MIRC-X, while the bottom panel corresponds to the K-band MYSTIC. The ROTIR method enforces a spherical shape to the star. The bright and dark spots on the star surface are interpreted similarly to those in Figure~\ref{Fig:oitools_images}. In the 2024 July and August epochs, the K-band images show three bright spots arranged in a V-shape on the star surface, while later epochs (September and October) reveal more bright spots, likely due to the time-varying dispersal of dust away from the star. The H-band images, with higher resolution, uncover more structures than the K-band images. The darker regions on the star surface correspond to cool, infalling material situated between these giant convection cells.} 
\label{Fig:ROTIR_images}
\end{figure*}

\begin{figure*}[h!]
\centering
\includegraphics[width=\textwidth]{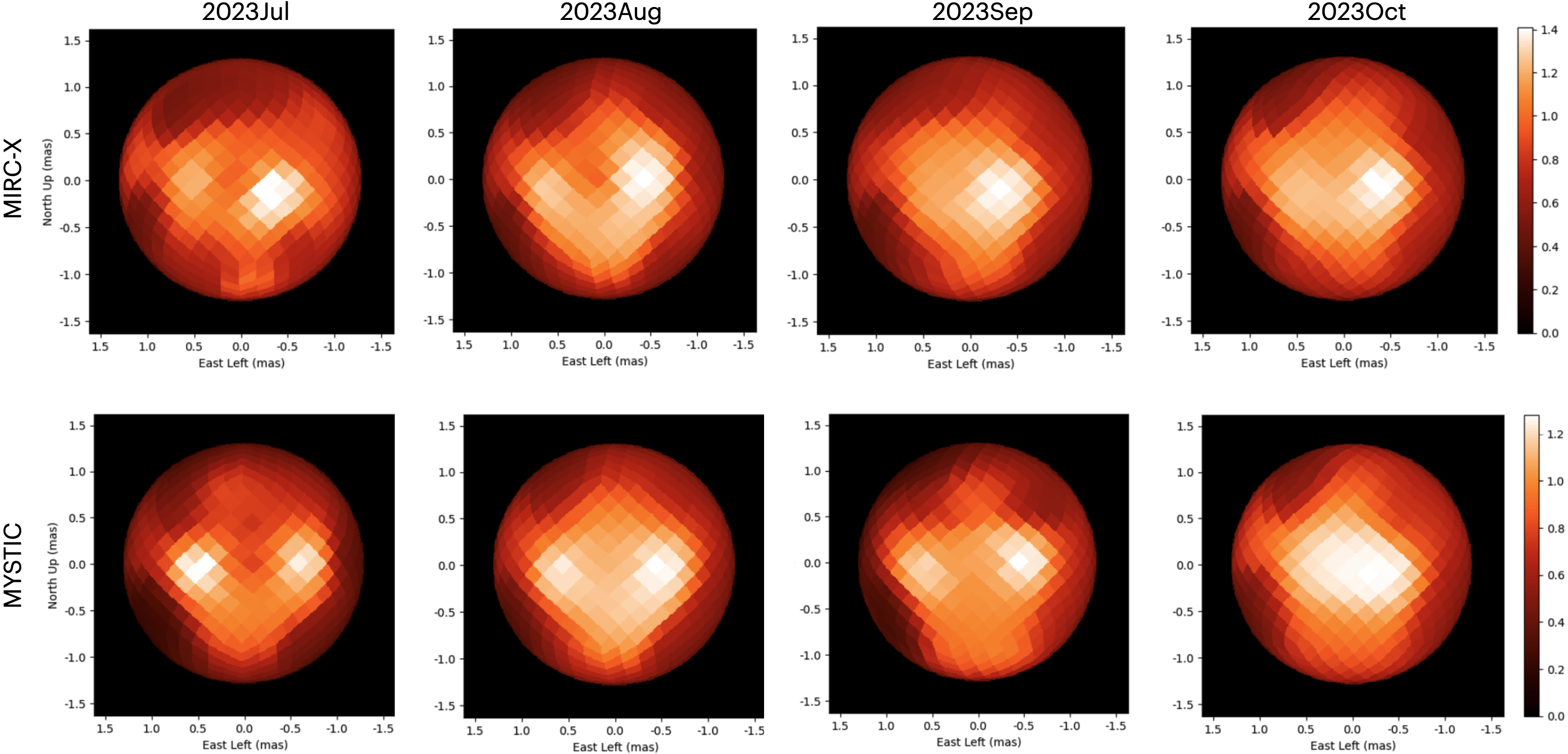}
\caption{The SURFING reconstructed images of RW~Cep for epochs 2023 July, August, September and October. The top panel corresponds to the H-band MIRC-X, while the bottom panel corresponds to the K-band MYSTIC. These images are interpreted in the same way as those in Figure~\ref{Fig:ROTIR_images}. However, the SURFING method shows that the bright and dark spots appear larger, even though the pixel scales are similar to those in ROTIR.}
\label{Fig:SURFING_images}
\end{figure*}

\section{Image Reconstructions Using Synthetic Data to Assess the Effects of Incomplete (u,v)-Coverage}\label{Sec:Synthetic}

Figures~\ref{Fig:oitools_simulated} and \ref{Fig:rotir_simulated} depict the synthetic image reconstructions intended to assess the fidelity of the image reconstructions shown in Figures~\ref{Fig:oitools_images} and \ref{Fig:ROTIR_images}. These synthetic observations simulate a featureless, limb-darkened disk star with a diameter of $\theta=2.6$ mas, replicating the same $(u,v)$ coverage and signal-to-noise ratio. The resulting images exhibit no significant features, suggesting that there are no major artifacts in the image reconstructions resulting from gaps in the $(u,v)$ coverage.

\begin{figure*}[h]
\centering
\includegraphics[width=\textwidth]{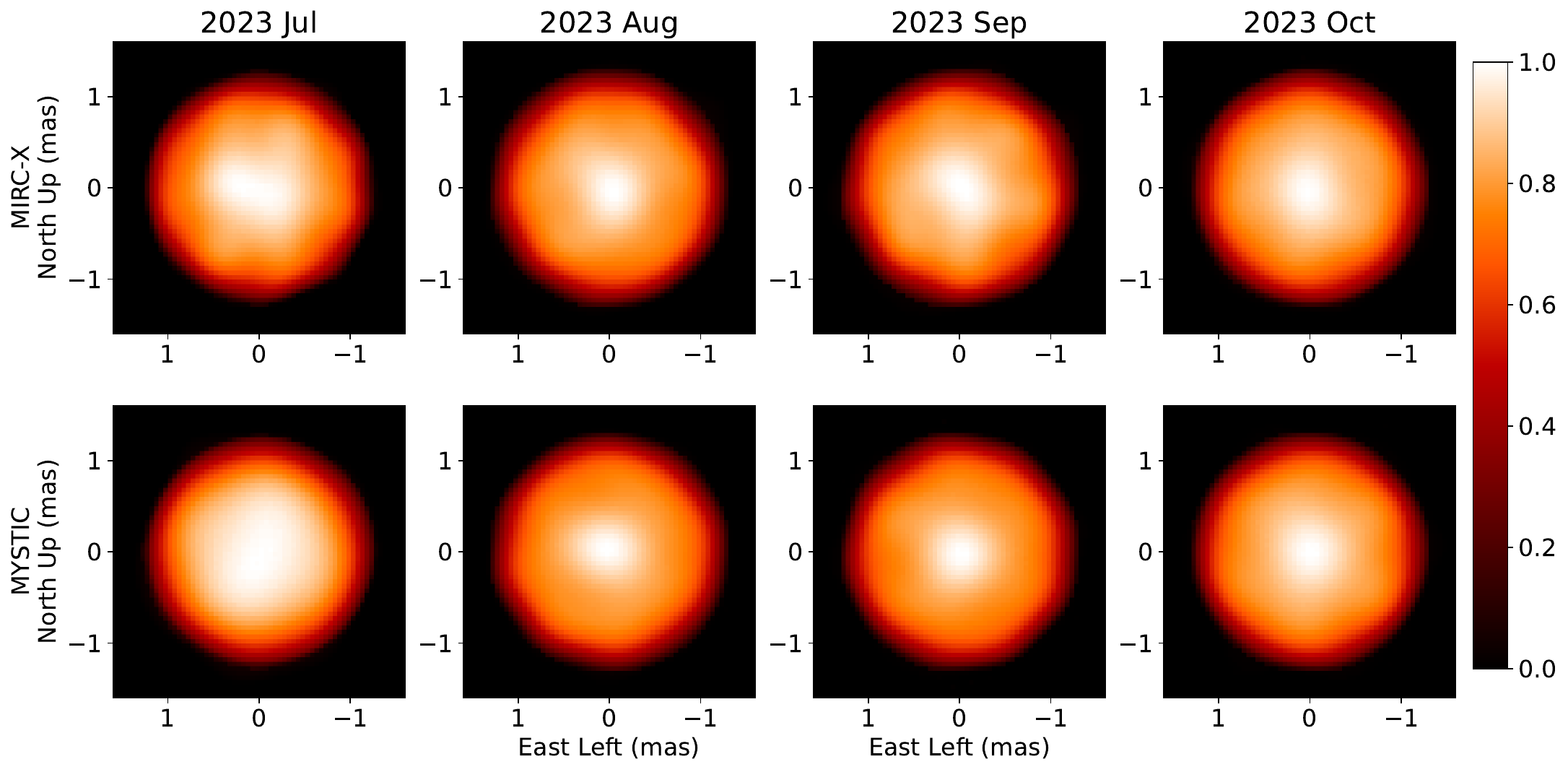}
\caption{Synthetic images reconstructed using the OITOOLS software. The synthetic observations,  replicate the same $(u,v)$ coverage and SNR as the real observations. 
}
\label{Fig:oitools_simulated}
\end{figure*}

\begin{figure*}[h]
\centering
\includegraphics[width=\textwidth]{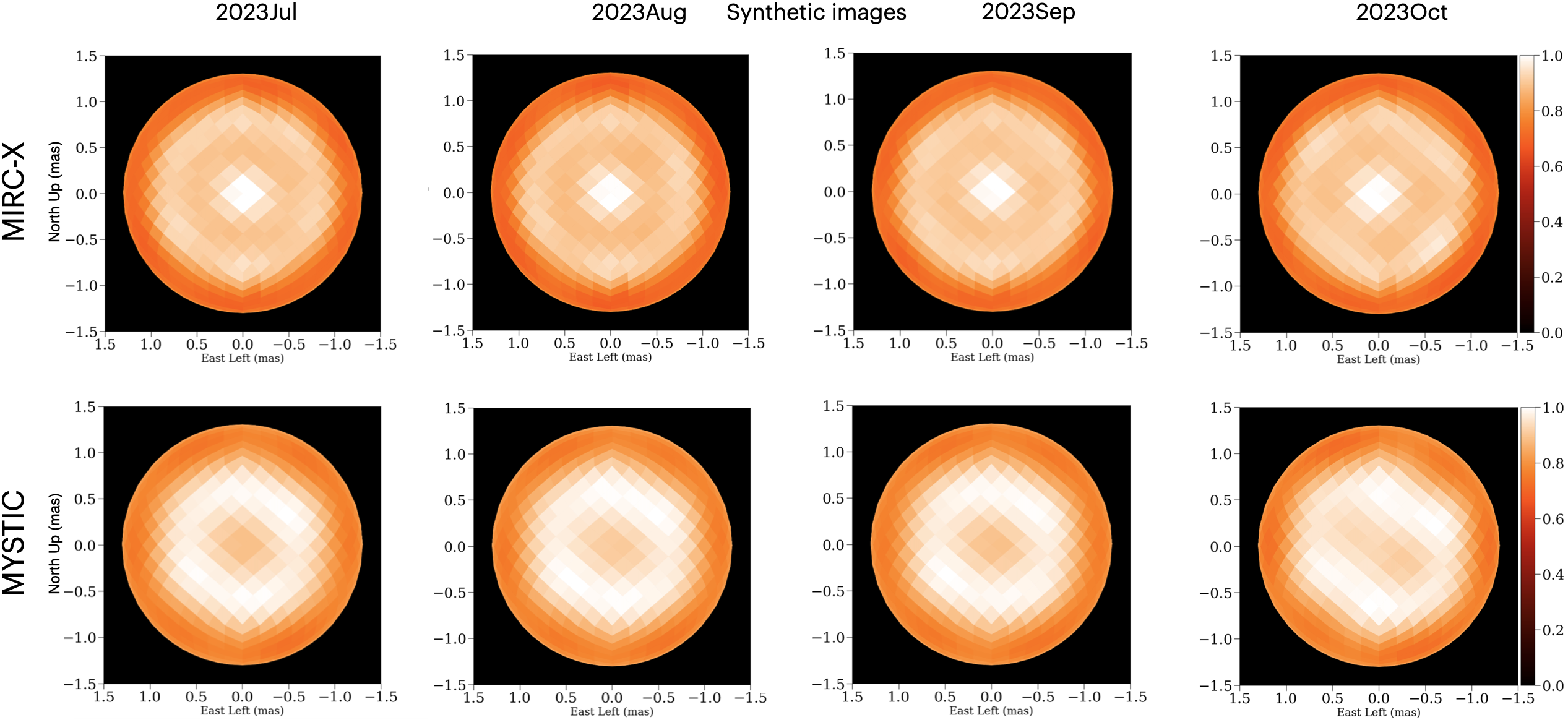}
\caption{These ROTIR images are reconstructed from the synthetic observations, which replicate the same $(u,v)$ coverage and SNR as the real observations.  
\label{Fig:rotir_simulated}}
\end{figure*}

\bibliography{ms3.bib}{}
\bibliographystyle{aasjournal}

\end{document}